\documentclass[twocolumn]{aastex701}
\shorttitle{AL for Planet Habitability}
\shortauthors{R.\,I.\,El-Kholy \& Z.\,M.\,Hayman}
\received{February 27, 2026}
\revised{\today}
\accepted{\today}
\submitjournal{PASP}

\usepackage{amssymb, amsmath, soul}

\begin{document}

\title{Active Learning for Planet Habitability Classification under Extreme Class Imbalance}

\correspondingauthor{R.~I.~El-Kholy}
\email{relkholy@sci.cu.edu.eg, reham.elkholy@cu.edu.eg}

\author[0000-0001-6337-1384]{R.~I.~El-Kholy}
\affiliation{Department of Astronomy, Space Science, and Meteorology \\
	Faculty of Science, Cairo University \\
	1 Gamaa Street, Giza 12613, Egypt}
\email{relkholy@sci.cu.edu.eg}

\author[0009-0009-4770-5626]{Z.~M.~Hayman}
\affiliation{Department of Astronomy, Space Science, and Meteorology \\
	Faculty of Science, Cairo University \\
	1 Gamaa Street, Giza 12613, Egypt} 
\email{zmhayman@cu.edu.eg}

\begin{abstract}
The increasing size and heterogeneity of exoplanet catalogs have made systematic habitability assessment challenging, particularly given the extreme scarcity of potentially habitable planets and the evolving nature of their labels. In this context, labels from the Habitable Worlds Catalog (HWC) should be interpreted as heuristic proxies rather than definitive physical classifications. In this study, we explore the use of pool-based active learning to improve the efficiency of learning such heuristic classifications under realistic observational constraints. We construct a unified dataset from the Habitable World Catalog and the NASA Exoplanet Archive and formulate habitability assessment as a binary classification problem. A supervised baseline based on gradient-boosted decision trees is established and optimized for recall in order to prioritize the identification of rare potentially habitable planets. This model is then embedded within an active learning framework, where uncertainty-based margin sampling is compared against random querying across multiple runs and labeling budgets. We find that active learning substantially reduces the number of labeled instances required to approach supervised performance, demonstrating clear gains in label efficiency. We further compare model predictions with an independent, physically motivated proxy habitability index, finding partial but non-trivial agreement. To connect these results to a practical astronomical use case, we aggregate predictions from independently trained active-learning models into an ensemble and use the resulting mean probabilities and uncertainties to rank planets originally labeled as non-habitable. This procedure identifies a single robust candidate for further study, supporting conservative prioritization without implying physical reclassification. Our results indicate that active learning provides a principled framework for label-efficient prioritization in exoplanet studies under class imbalance and limited data.
\end{abstract}



\section{Introduction}
\label{sec:intro}
The increasing scale and complexity of modern astronomical surveys have intensified the need for machine learning (ML) methods capable of extracting reliable scientific insights from large, heterogeneous datasets. In many astronomical applications, the primary bottleneck is not data availability but the scarcity of reliable labels, which often require costly follow-up observations or expert analysis \citep{Richards2011}. Active learning (AL) addresses this challenge by enabling models to iteratively select the most informative unlabeled instances for annotation, thereby improving predictive performance while minimizing labeling effort \citep{Ishida2018, Skoda2020}.

AL differs fundamentally from conventional supervised learning by replacing static or random training set construction with a sequential querying process. At each iteration, the learner identifies data points whose labels are expected to yield the greatest improvement to the model, querying an oracle for their true labels. This approach has proven effective in settings where labeled data are limited, expensive, or unevenly distributed, making it particularly relevant for many astronomical problems \citep{Solorio2005, Richards2011, Masters2015, Hoyle2016}.

In astronomy, AL has been successfully applied to a range of tasks, including variable star classification \citep{Richards2011}, supernova photometric classification \citep{Ishida2018, Ishida2021, Kennamer2020, Leoni2022}, galaxy morphology classification \citep{Walmsley2019}, stellar parameters inference \citep{ElKholy2025}, and anomaly detection in large survey datasets \citep{Ishida2021, Lochner2021}. These studies demonstrate that AL can significantly improve model performance or reduce labeling requirements when compared to random sampling strategies, particularly in scenarios characterized by high data volume and limited labeling resources.

However, prior work \citep{Zhang2018, Yu2019, Quan2021} has also highlighted several challenges associated with applying AL in astronomical contexts. A recurring issue is severe class imbalance, where scientifically interesting objects constitute a small fraction of the overall dataset. Under such conditions, common uncertainty-based querying strategies may perform suboptimally, motivating careful evaluation of AL behavior in imbalanced regimes.

Despite the breadth of AL applications in astronomy, its use in the context of exoplanet habitability assessment remains notably unexplored. Existing ML studies on planet habitability focus primarily on supervised classification \citep{Saha2018, Saha2020, Basak2021}, regression-based habitability indices \citep{SchulzeMakuch2011, RodriguezMozos2017, Patel2024, Dhama2025}, or architectural comparisons \citep{Barnes2015}, without considering AL as a means of addressing label scarcity or prioritizing follow-up observations. This omission is striking given that habitability labels are inherently uncertain, sparse, costly to validate, and typically derived from heuristic criteria applied to highly imbalanced datasets.

This work addresses this gap by investigating the application of AL to the classification of proxy habitability labels. In particular, we treat labels derived from the Habitable Worlds Catalog (HWC) as heuristic indicators rather than definitive physical measurements of habitability. Building on established AL methodologies developed in astronomical and ML contexts, we evaluate whether uncertainty-driven instance selection can improve classification performance relative to random sampling under realistic data constraints. Rather than attempting to infer physical habitability directly, the objective is to assess whether AL can improve the efficiency of learning and reproducing such heuristic label definitions, and to what extent the resulting model behavior aligns with independent physically motivated proxies. We further examine the implications of AL for prioritizing candidate planets for further study. This prioritization is interpreted in a conservative, uncertainty-aware sense, and does not imply physical reclassification of planetary habitability. The methodology and experimental design are described in Section~\ref{sec:methods}, with results presented in Section~\ref{sec:res} and their implications discussed in Section~\ref{sec:disc}. We give our final conclusions in Section~\ref{sec:conc}. All data and source code used in this study are made available in a public GitHub repository\footnote{\url{https://github.com/rehamelkholy/ExoplanetAL}}.

\section{Data and Methods}
\label{sec:methods}
\subsection{Dataset and feature selection}
\label{subsec:data}
In this work, two publicly available data sources were used. The first is the HWC \citep{HWC2025}, which lists 70 potentially habitable worlds out of over 5000 known exoplanets. They are further divided into a conservative and optimistic sample. The conservative subset comprises planets with radii not exceeding 1.6 Earth radii or masses below 3 Earth masses, which are more likely to be rocky in composition. The optimistic subset extends to larger planets, potentially encompassing super-Earths, ocean worlds, or mini-Neptunes, which are correspondingly less likely to be rocky or to sustain surface liquid water. We used a version of the full catalog downloaded on 22 October 2025. It is important to note that the HWC classification is based on a set of heuristic criteria, including habitable zone location, planetary size, and derived indices such as the Earth Similarity Index (ESI) \citep{SchulzeMakuch2011}, and does not constitute a physically complete or self-consistent model of planetary habitability.

The HWC primarily draws its exoplanet parameters from the Planetary Systems Composite Parameters (PSCompPars) table of the NASA Exoplanet Archive\footnote{\url{https://exoplanetarchive.ipac.caltech.edu/index.html}}, which is the second source we used \citep{Christiansen2025}, also downloaded on 22 October 2025. However, the HWC is supplemented by additions and corrections from other literature sources. While this approach yields a more comprehensive collection of planetary properties, it also introduces potential inconsistencies arising from the aggregation of independent measurements. Consequently, the HWC is best regarded as a heuristic reference framework to guide further investigation rather than a definitive classification of planetary habitability. Because individual references do not necessarily report a complete set of stellar and planetary properties, the PSCompPars table itself consolidates measurements from multiple references into a single row per confirmed planet. This aims to maximize parameter completeness by selecting preferred values across sources, yielding a more filled-in representation of planetary and stellar properties. However, this increased completeness comes at the expense of strict self-consistency, as the reported parameters may originate from different studies employing distinct assumptions or methodologies. Accordingly, the PSCompPars table is well suited for statistical and ML applications that require broad feature coverage, but its contents should be interpreted with appropriate caution.

The downloaded version of the HWC contained 5599 exoplanets, while the PSCompPars table contained 6028 exoplanets. The two datasets were merged by cross-matching the planetary names. The resulting merged dataset contained 5576 exoplanets, including the 70 potentially habitable planets. In this work, the HWC labels are used as target variables for supervised learning and active learning experiments, and are therefore interpreted strictly as proxy labels reflecting the underlying heuristic definition rather than ground-truth indicators of physical habitability.

The feature selection process was designed to construct a physically meaningful and internally consistent representation of planetary, stellar, and system properties, while making maximal use of the information available across the two source catalogs. Following the cross-matching of the HWC and the PSCompPars table, all quantitative parameters from both sources were initially retained in a merged dataset to allow explicit evaluation of overlap and redundancy.

Because the two catalogs report many of the same physical quantities---such as orbital parameters, planetary size and mass estimates, and stellar properties---the merged dataset contains multiple column pairs describing identical or closely related attributes. As a first step in resolving this redundancy, the degree of agreement between corresponding columns from the two sources was assessed using pairwise correlation analysis. This analysis confirmed that overlapping parameters from HWC and PSCompPars are strongly correlated, indicating that they encode the same underlying physical information rather than independent measurements.

After establishing this equivalence, redundancy was resolved by comparing the completeness of each overlapping column pair. For each physical quantity represented in both catalogs, the column exhibiting superior overall completeness was selected as the primary representation and retained for subsequent analysis. Following this selection, the redundant column from the other source was used exclusively to supplement missing values in the retained column wherever possible. This ensured that each physical quantity is represented by a single unified column in the final dataset. Once supplementation was complete, the redundant columns were removed. This procedure prioritizes feature completeness and coverage, which is advantageous for ML applications, but may introduce implicit inconsistencies due to the heterogeneous origin of the underlying measurements. These effects are considered when interpreting model outputs.

\begin{figure*}
	\centering
	\includegraphics[width=\textwidth]{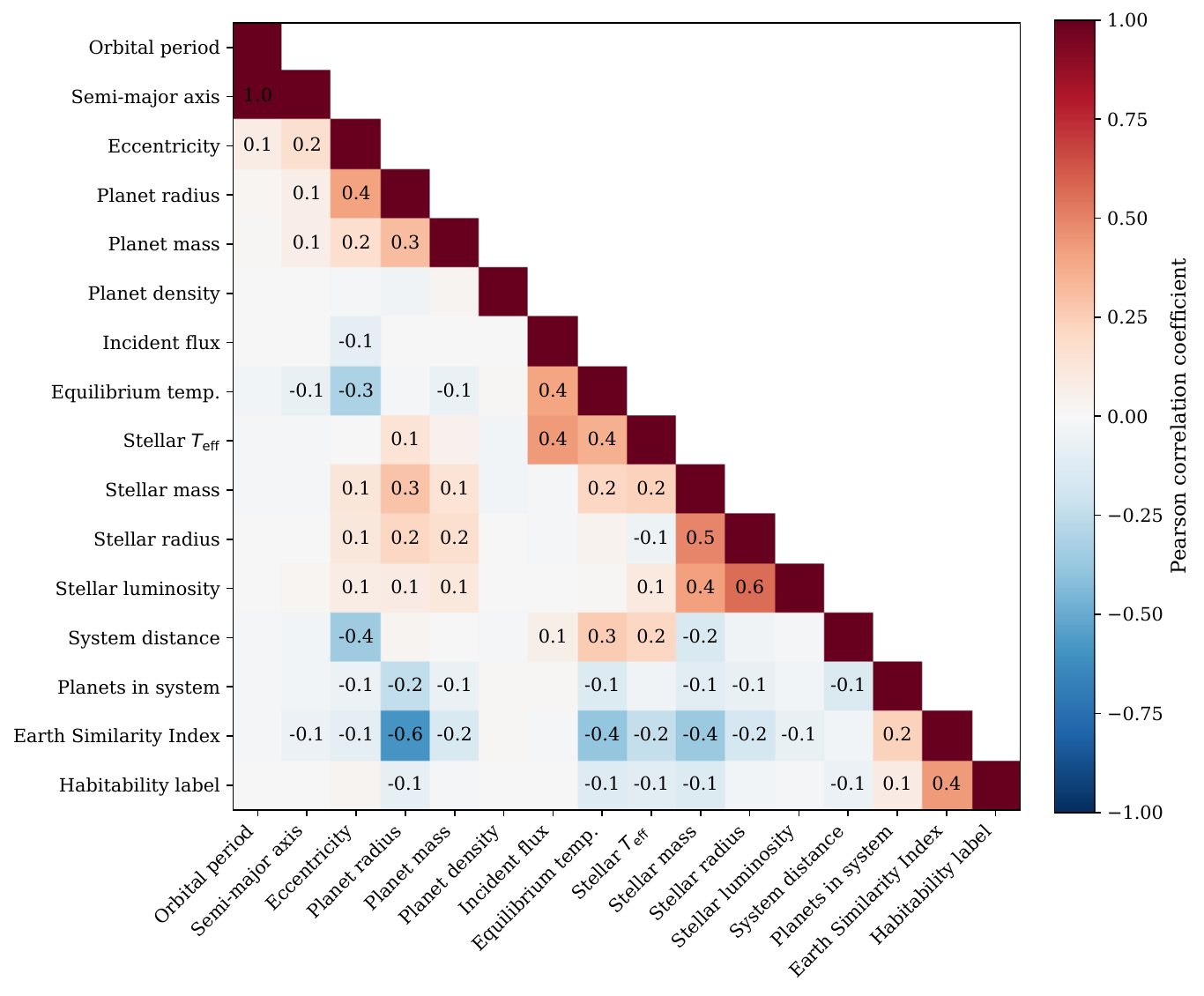}
	\caption{Lower-triangular Pearson correlation matrix of the final feature set used in this study. Each cell shows the Pearson correlation coefficient between a pair of planetary or stellar properties, with the color scale ranging from $-1$ (strong negative correlation) to $+1$ (strong positive correlation). The diagonal elements represent self-correlations. To improve readability, only non-zero correlations are annotated.}\label{fig:corr}
\end{figure*}

The resulting feature set was then examined through exploratory analysis conducted entirely on this final, unified set of features. Pairwise correlations among retained features were inspected to characterize dependencies arising from known physical relationships. The resulting visualization is shown in Figure~\ref{fig:corr}. It reveals several physically expected dependencies. Strong positive correlations are observed among stellar properties, particularly between stellar radius and stellar luminosity, reflecting underlying stellar structure relations. Planetary radius and planetary mass also show a moderate positive correlation, consistent with known mass-radius trends for exoplanets. Incident stellar flux and planetary equilibrium temperature are moderately correlated, as expected from their shared dependence on stellar luminosity and orbital separation. A notable moderate negative correlation is observed between ESI and planetary radius, indicating that larger planets tend to score lower on Earth-likeness metrics. Because ESI is itself a composite index derived from multiple planetary properties, its inclusion as a feature introduces an element of built-in heuristic structure that overlaps with the HWC labeling criteria. This is taken into account when interpreting feature importance and model behavior in subsequent sections. Importantly, no pair of retained features exhibits near-perfect correlation outside of these physically motivated relationships, suggesting that the final feature set does not suffer from severe multicolinearity. This supports the suitability of the selected variables for subsequent ML and AL analyses.

\begin{figure*}
	\centering
	\includegraphics[width=\textwidth, trim= 180 0 50 0, clip]{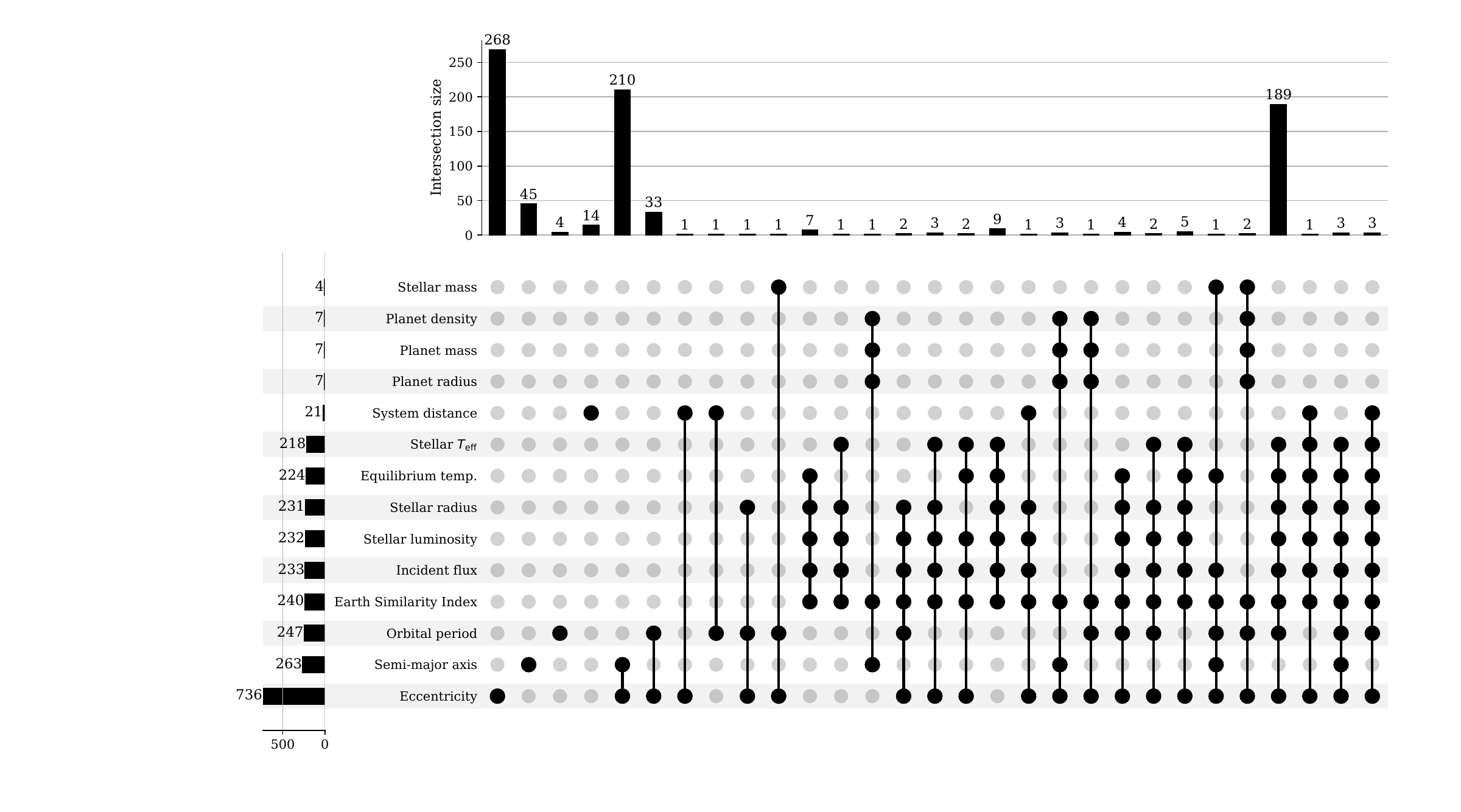}
	\caption{UpSet plot showing missing-value patterns across the final feature set. The horizontal bars on the left indicate the total number of missing values for each feature. Each column in the central matrix represents a specific combination of features that are simultaneously missing, marked by filled black circles connected by vertical lines. The vertical bars above the matrix show the number of data instances exhibiting each missingness pattern (intersection size). For example, the tallest bar on the far left corresponds to instances where only orbital eccentricity is missing, while all other features are present. In contrast, columns with multiple connected black circles indicate instances where several stellar and planetary properties are missing together.}\label{fig:missing}
\end{figure*}

The structure of missing values in the final dataset is summarized in Figure~\ref{fig:missing} using an UpSet \citep{Lex2014} representation. The plot reveals that missingness is not uniformly distributed across features, but instead exhibits clear patterns. Orbital eccentricity accounts for the largest share of missing values, frequently occurring as the sole missing parameter in otherwise complete records. In contrast, stellar luminosity, stellar radius, incident stellar flux, and planetary mass, radius, and density show comparatively low missingness. These structured missingness patterns indicate that absent values are primarily driven by observational and derivational limitations rather than random omission. Missing values are addressed through a dedicated preprocessing procedure described in Section~\ref{subsec:preprocessing}, where the impact of imputation and associated assumptions is discussed in detail.

\begin{figure*}
	\centering
	\includegraphics[width=\textwidth, trim= 0 0 0 0, clip]{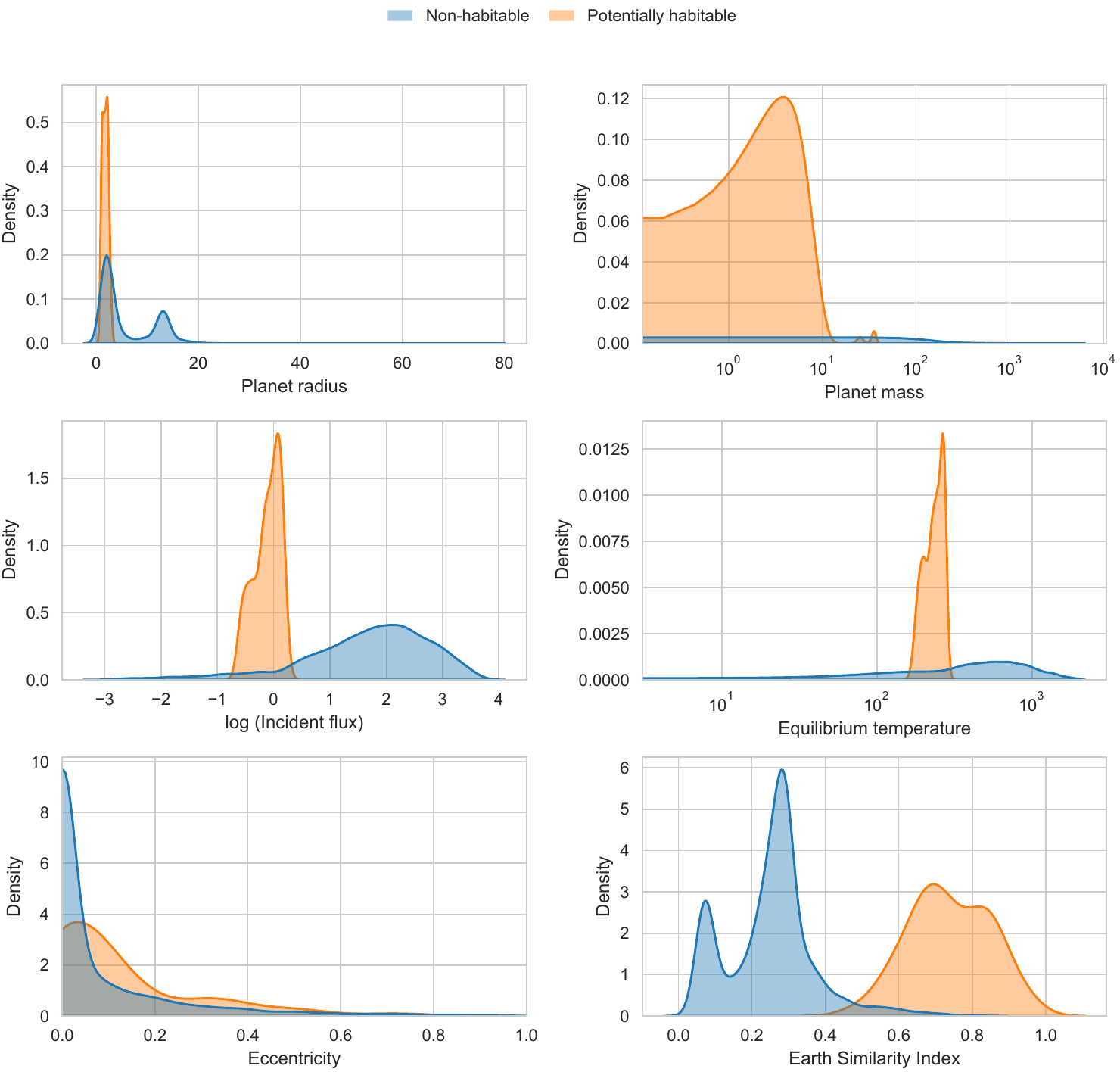}
	\caption{Class-conditional distributions of selected planetary and stellar parameters in the final dataset. Kernel density estimates are shown for planet radius, orbital eccentricity, and ESI; and in log scale for planet mass, incident stellar flux, and equilibrium temperature. The figure illustrates the substantial overlap between the two classes across most individual parameters, as well as systematic shifts in location and spread for several features, motivating the need for multivariate classification.}\label{fig:features}
\end{figure*}

Figure~\ref{fig:features} highlights the degree to which potentially habitable and non-habitable planets overlap in marginal feature space. The subset of features shown was selected to provide a representative cross-section of the physical parameters used in the classification task, including planetary size (radius, mass), irradiation conditions (incident flux, equilibrium temperature), orbital properties (eccentricity), and composite habitability metrics (ESI). These variables are commonly used in habitability assessments and capture the primary structure underlying the adopted heuristic labeling scheme. These distributions are shown individually to illustrate the degree of class overlap in marginal feature space, providing a qualitative assessment of feature separability and highlighting the limitations of relying on any single parameter for classification. While potentially habitable planets tend to cluster toward smaller radii and lower masses, the distributions remain broad and partially overlapping, reflecting both observational uncertainties and the inclusive definition of habitability adopted by the HWC. Incident stellar flux and equilibrium temperature exhibit clearer class separation when viewed on logarithmic scales, with potentially habitable planets concentrated near Earth-like fluxes and moderate equilibrium temperatures, whereas non-habitable planets span a much wider dynamic range. However, even in these parameters, no sharp decision boundary exists. Orbital eccentricity shows strong skewness toward low values in both classes, with potentially habitable planets slightly more  concentrated at near-circular orbits, consistent with expectations for long-term climate stability. The ESI provides the most visually distinct separation, by construction aggregating multiple planetary properties into a single heuristic score that is closely aligned with the labeling criteria, yet still exhibits overlap between classes. Overall, these class-conditional distributions demonstrate that habitability cannot be reliably inferred from univariate thresholds alone. Instead, the substantial overlap across most features motivates the use of multivariate ML models and, in particular AL strategies capable of efficiently identifying informative regions of parameter space where class distinctions are most ambiguous. Together, the final features provide a compact yet physically interpretable description of each planetary system, balancing completeness and redundancy while preserving all dimensions relevant to habitability assessment.

\subsection{Problem formulation}
\label{subsec:problem}
We formulate the task as a binary classification problem, where each confirmed exoplanet is assigned a label indicating whether it is potentially habitable or non-habitable based on current catalog assessments. Given a feature vector containing stellar, planetary, and system-level properties for a planet, the objective is to learn a classifier that maps the vector onto a binary output indicating whether the planet is classified as potentially habitable or non-habitable. In this formulation, the model learns to approximate the mapping defined by the HWC labeling scheme, rather than to infer physical habitability directly. This reflects a practical observational setting in which habitability is treated as a prioritization signal rather than a definitive physical state. The classifier is therefore designed to identify planets that merit further study under limited observational resources, rather than to determine true biological habitability.

Positive labels are derived from the HWC, which identifies planets that orbit within the stellar habitable zone and satisfy broad constraints on planetary size and mass, aiming to be inclusive while acknowledging substantial astrophysical uncertainties \citep{HWC2025}. As mentioned in Section~\ref{subsec:data}, the catalog subdivides potentially habitable planets into conservative and optimistic samples. In this study, both samples are labeled as potentially habitable. This choice reflects the study's emphasis on accelerating candidate discovery rather than enforcing strict geophysical definitions. Treating the two samples jointly increased the number of positive instances and aligns with the goal of identifying planets worthy of follow-up observations under uncertainty.

Importantly, these labels should not be interpreted as ground truth indicators of actual habitability. As emphasized in the HWC documentation \citep{HWC2025}, the criteria used to define habitability rely on simplified models (e.g., \citet{SchulzeMakuch2011, Kopparapu2014}), incomplete measurements, and assumptions about atmospheric composition and planetary structure. Consequently, some positively-labeled planets may ultimately prove uninhabitable, while others currently labeled as non-inhabitable could later be revised as new observations and improved models become available. The labels therefore represent heuristic, literature-based assessments rather than definitive physical classifications. In later sections, we explicitly assess the extent to which model predictions align with independent, physically motivated proxy metrics. This uncertainty motivates the use of ML models and AL strategies that can operate effectively under label noise and class imbalance.


\subsection{Preprocessing and missing-data handling}
\label{subsec:preprocessing}
The merged dataset resulting from the feature selection stage (Section~\ref{subsec:data}) required several preprocessing steps to address missing values and physical consistency prior to model training and AL. These steps were designed to minimize information loss while preserving plausibility and avoiding the introduction of bias into the habitability labels. Throughout this process, all input quantities are treated as deterministic point estimates, and observational uncertainties are not explicitly propagated through the feature construction or modeling pipeline. The implications of this assumption are evaluated through sensitivity analyses presented in Section~\ref{subsec:ML}.

As a first step, we attempted to reduce missingness using deterministic physical relationships between planetary and stellar parameters wherever possible. This approach was preferred over statistical imputation when reliable formulae could be applied, as it preserves known astrophysical constraints. Only two such derivations were possible. For planets with a known orbital period and host stellar mass, missing values of the orbital semi-major axis were computed using Kepler's third law,
\begin{equation}
	a = \left(\frac{G M_{\star} P^2}{4 \pi^2}\right)^{1/3},
\end{equation}
where $a$ is the semi-major axis, $P$ is the orbital period, $M_\star$ is the stellar mass, and $G$ is the gravitational constant. Planetary mass was neglected relative to the stellar mass, consistent with standard practice for exoplanetary systems. This relation allowed the recovery of 259 additional semi-major axis values, substantially reducing missingness in this parameter. Uncertainties in stellar mass propagate directly into the derived semi-major axis; however, these uncertainties are not explicitly modeled and are instead assessed through downstream sensitivity tests. For stellar systems with known stellar luminosity and effective temperature but missing stellar radius, the radius was derived using the Stefan-Boltzmann relation,
\begin{equation}
	L_\star = 4 \pi R^2_\star \sigma T^4_{\text{eff}},
\end{equation}
which can be rearranged as 
\begin{equation}
	R_\star = \left(\frac{L_\star}{4 \pi \sigma T^4_{\text{eff}}}\right)^{1/2},
\end{equation}
where $L_\star$, $R_\star$, and $T_\text{eff}$ are stellar luminosity, radius, and effective temperature, respectively, and $\sigma$ is the Stefan-Boltzmann constant. This relation resulted in one additional valid stellar radius value. After applying these relations, all derived values were validated against physical bounds.

Following physics-based completion, missing values were reassessed for both the full dataset and the subset of potentially habitable planets. Among the 70 planets labeled as potentially habitable, missing values were found only in orbital eccentricity (10 entries). All other features were complete for this class. Across the full dataset, the remaining missing values were confined to a small number of parameters, including planetary density, incident flux, equilibrium temperature, stellar radius, stellar luminosity, and a small number of semi-major axis entries. Given the large size of the non-habitable class, rows with missing values in these features were removed, while rows missing only eccentricity were retained for imputation. This filtering resulted in a dataset of 5281 planets, preserving all potentially habitable instances while ensuring completeness for the remaining features.

Before imputation, all features were subjected to sanity checks to identify invalid or extreme values. These included enforcing non-negativity for quantities such as mass, radius, flux, and semi-major axis; and constraining orbital eccentricity to the physical range $[0,1]$. No systematic anomalies were found beyond those already addressed through row removal or subsequent imputation.

Orbital eccentricity was imputed using a supervised regression approach trained on planets with complete feature vectors. From the 4803 planets with observed eccentricity, the data were split into an 80\% training set and a 20\% test set. A standard scaler was fitted on the training data and applied consistently to the test set and the rows requiring imputation. Two regressors were evaluated: a random forest regressor and a gradient boosting regressor. Model selection was based on mean absolute error (MAE) computed on the held-out test set. Gradient boosting achieved the lowest error and was therefore selected for imputation. To quantify uncertainty, seven imputation iterations were performed using bootstrap resampling of the training set while keeping hyperparameters fixed. For each missing entry, the final imputed value was computed as the mean of the seven predictions, and uncertainty was estimated as their standard deviation. All imputed values were clipped to the physical interval $[0,1]$, and additionally constrained to lie within $[\min_{\text{neighbor}} - 3 \sigma, \max_{\text{neighbor}} + 3 \sigma]$, where neighboring values were defined in feature space.

\begin{figure}
	\centering
	
	\begin{minipage}{0.95\columnwidth}
		\centering
		\includegraphics[height=0.23\textheight]{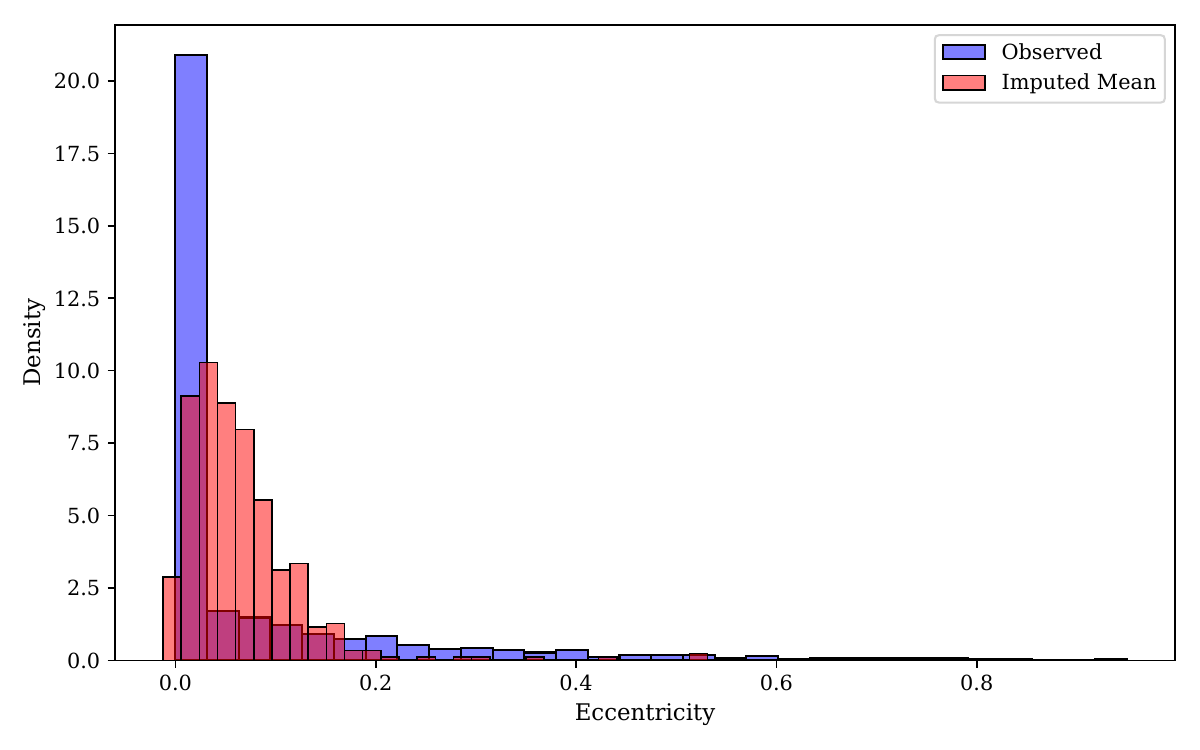}\\
		\centerline{\small (a)}
	\end{minipage}
	\hfill
	\begin{minipage}{0.95\columnwidth}
		\centering
		\includegraphics[height=0.23\textheight]{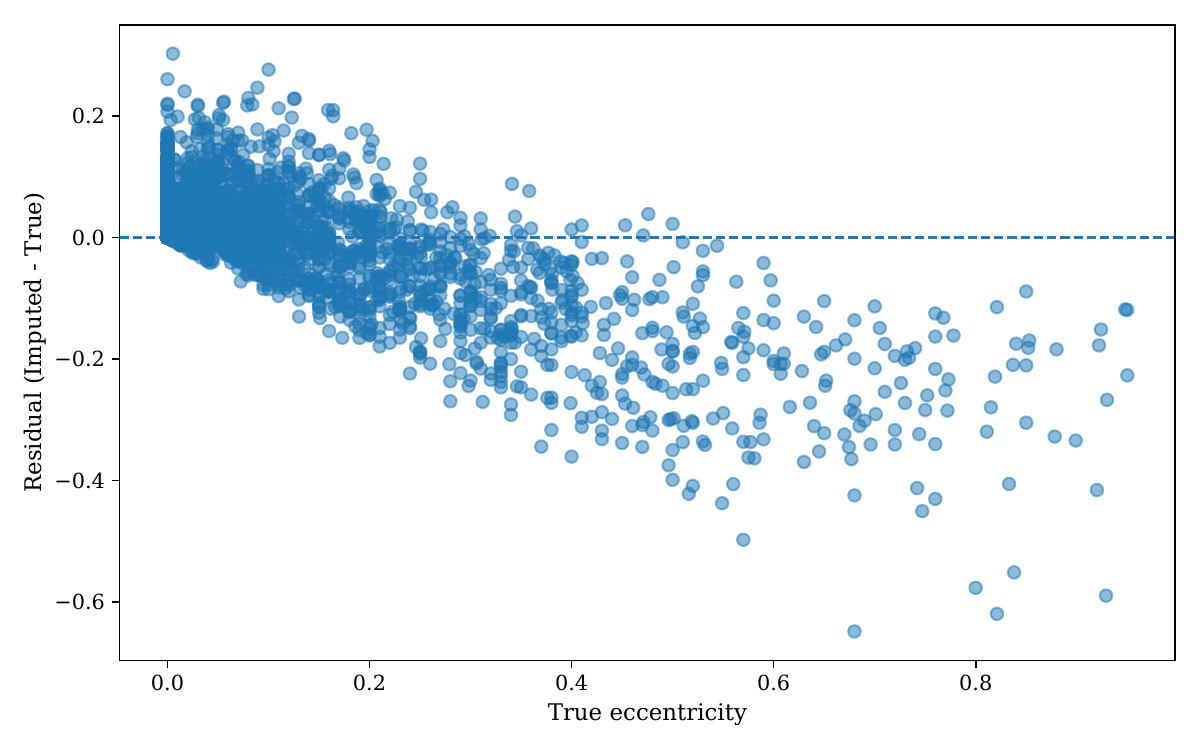}\\
		{\small (b)}
	\end{minipage}
	\hfill
	\begin{minipage}{0.95\columnwidth}
		\centering
		\includegraphics[height=0.23\textheight]{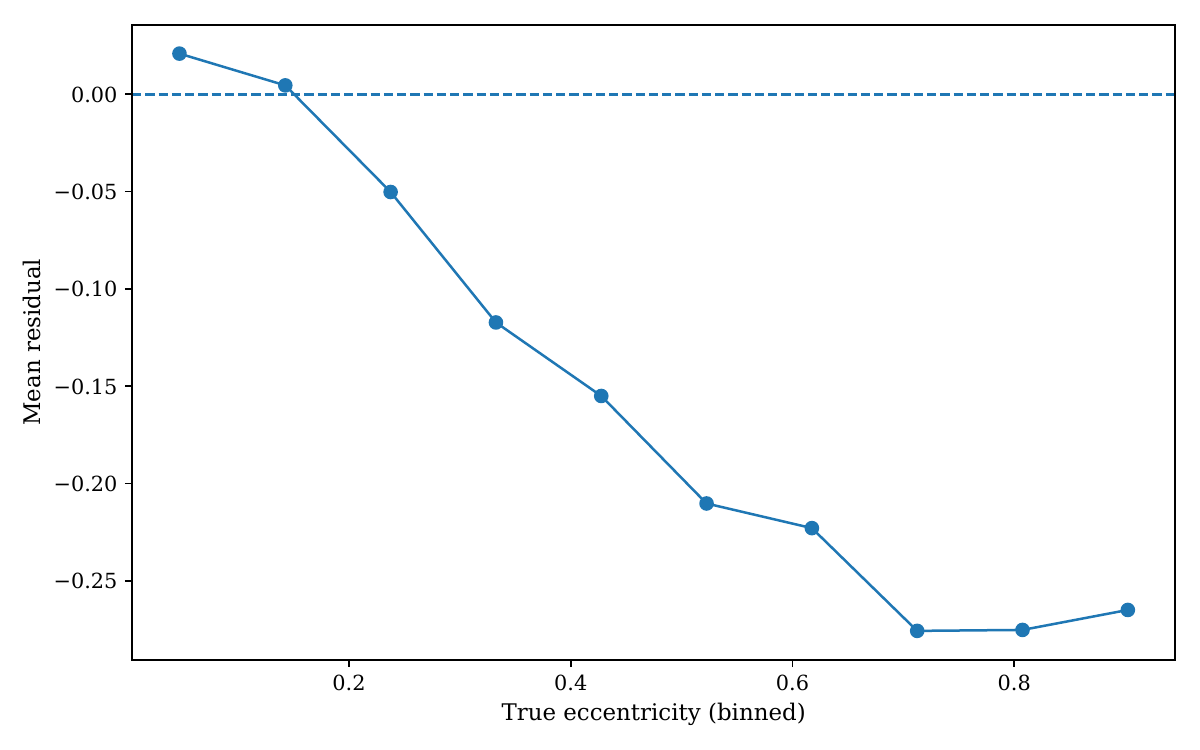}\\
		{\small (c)}
	\end{minipage}
	\caption{Diagnostics of orbital eccentricity imputation. Panel (a) compares observed and imputed eccentricity distributions; panel (b) shows residuals as a function of true eccentricity; and panel (c) shows the mean residual in bins of true eccentricity, highlighting a systematic bias toward under estimating at higher eccentricities.}
	\label{fig:imputation}
\end{figure}

Figure~\ref{fig:imputation} summarizes the diagnostics used to assess the behavior of the orbital eccentricity imputation. Panel (a) compares the empirical distribution of observed eccentricities with the distribution of imputed values. The close agreement between the two indicates that the imputation procedure preserves the overall shape of the eccentricity distribution and does not introduce artificial modes or inflate the high-eccentricity tail. Panel (b) shows residuals (imputed minus true eccentricity) as a function of the true value, revealing increasing dispersion and a systematic tendency toward underestimation at higher eccentricities. Panel (c) shows the mean residual as a function of binned true eccentricity, demonstrating a monotonic trend toward increasingly negative bias with increasing eccentricity. These results indicate that the imputation procedure is distributionally consistent but exhibits structured bias, particularly in sparsely sampled regions of parameter space. This behavior is consistent with both the reduced density of high eccentricity training examples and the large observational uncertainties typically associated with eccentricity measurements.

After preprocessing and imputation, the final dataset consisted of 5821 planets with complete feature vectors, including all 70 potentially habitable planets. This deterministic representation of the dataset was used unchanged in all subsequent baseline modeling and AL experiments.

\subsection{Evaluation metrics}
\label{subsec:metrics}
The planet habitability prediction task considered in this work is formulated as a highly imbalanced binary classification problem, with potentially habitable planets constituting a small minority of the overall sample. As a consequence, standard accuracy alone is not a meaningful indicator of model performance, since a trivial classifier predicting all planets as non-habitable would achieve a deceptively high accuracy. Instead, a set of complementary evaluation metrics \citep{Acquaviva2023} was employed to assess different aspects of classifier behavior, with particular emphasis on sensitivity to the minority class.

Recall, also referred to as sensitivity or true positive rate, is defined as
\begin{equation}
	\text{Recall} = \dfrac{\text{TP}}{\text{TP} + \text{FN}},
\end{equation}
where TP and FN denote the number of true positives and false negatives, respectively. Recall quantifies the fraction of planets labeled as potentially habitable that are correctly identified by the model. In the present context, recall is of primary importance, as failing to identify a potentially habitable planet represents a more severe error than incorrectly flagging a non-habitable one. Precision measures the reliability of positive predictions and is given by
\begin{equation}
	\text{Precision} = \dfrac{\text{TP}}{\text{TP + \text{FP}}},
\end{equation}
where FP denotes the number of false positives. Precision complements recall by indicating the degree of contamination in the set of planets predicted as potentially habitable. The F1-score is defined as the harmonic mean of precision and recall,
\begin{equation}
	\text{F1} = 2 \times \dfrac{\text{Precision} \times \text{Recall}}{\text{Precision} + \text{Recall}},
\end{equation}
and provides a single scalar summary of performance when a balance between sensitivity and reliability is desired.

To account explicitly for class imbalance, balanced accuracy was also computed. Balanced accuracy is defined as the average of the recall obtained on each class,
\begin{equation}
	\text{Balanced Accuracy} = \dfrac{1}{2} \left(\dfrac{\text{TP}}{\text{TP} + \text{FN}} + \dfrac{\text{TN}}{\text{TN} + \text{FP}}\right),
\end{equation}
where TN denotes true negatives. This metric penalizes classifiers that perform well on the majority class but poorly on the minority class. Finally, the area under the receiver operating characteristic curve (AUROC) \citep{Fawcett2006} was used to assess the ranking quality of probabilistic predictions independently of any fixed classification threshold. The AUROC measures the probability that a randomly chosen potentially habitable planet is assigned a higher predicted probability than a randomly chosen non-habitable one.

Across all experiments, recall was treated as the primary optimization metric, reflecting the objective of minimizing missed candidates under the adopted labeling scheme. The remaining metrics were used to provide a more complete characterization of classifier performance and to facilitate comparisons between different modeling approaches. This choice is consistent with the prioritization-oriented interpretation of the task, in which the goal is to identify candidates for further study rather than to establish physical habitability.

\subsection{Supervised learning baseline}
\label{subsec:ML}
To establish a strong reference point for subsequent AL experiments, we first evaluated a set of supervised learning models trained on the fully preprocessed dataset described in Section~\ref{subsec:preprocessing}. These baselines serve two purposes: they quantify the achievable performance under standard passive learning, and they provide a fixed classifier architecture to be used within the AL framework introduced in later sections. In this context, the supervised models are trained to reproduce the mapping defined by the HWC labeling scheme, and their performance should therefore be interpreted with respect to this heuristic classification rather than as a direct measure of physical habitability.

The dataset was divided into training and testing subsets using an $80/20$ split, stratified by the habitability label to preserve the strong class imbalance between potentially habitable and non-habitable planets. Stratification ensures that both subsets contain comparable proportions of positive and negative instances, which is essential for meaningful evaluation under imbalance. Three supervised classifiers were considered:
\begin{enumerate}
	\item Random Forest (RF) \citep{Breiman2001}, an ensemble of decision trees trained on bootstrap samples with randomized feature selection, known for robustness to noise and nonlinear feature interactions.
	\item Extreme Gradient Boosting (XGBoost) \citep{Chen2016}, a gradient-boosted tree ensemble that sequentially fits weak learners to correct previous errors, incorporating regularization and subsampling to control overfitting.
	\item Multilayer Perceptron (MLP) \citep{Bourlard1994}, a feedforward neural network trained via backpropagation, representing a parametric, non-tree-based alternative.
\end{enumerate}
These models were selected to cover a range of inductive biases and learning paradigms commonly used in astrophysical classification tasks (e.g., \citet{Fluke2019, Andres2022, Agarwal2023, Zeraatgari2023, Bhavanam2024}). All performance evaluation relied on the metrics defined in Section~\ref{subsec:metrics}, with recall treated as the primary criterion due to the objective of minimizing missed candidates under the adopted labeling scheme.

To obtain unbiased estimates of generalization performance while tuning hyperparameters, a nested cross-validation strategy was employed. Nested cross-validation separates hyperparameter optimization from performance estimation and is particularly important when working with limited positive samples. An outer cross-validation loop consisting of five stratified folds repeated five times was used to assess model performance. Within each outer training fold, an inner stratified three-fold cross-validation loop was used to perform grid-based hyperparameter optimization \citep{Bergstra2012}. For each candidate model, a preprocessing-and-classification pipeline was constructed, consisting of a robust feature scaler followed by the classifier. Robust scaling \citep{Huber1981} was chosen to mitigate the influence of outliers and heavy-tailed feature distributions, which are common in exoplanetary and stellar parameters. Hyperparameter grids were explored independently for each model, and model refitting within the inner loop was performed using recall as the selection metric. For each outer fold, the best-performing estimator identified by the inner loop was evaluated on the held-out validation data, and performance metrics were recorded.

Performance metrics obtained from the outer cross-validation folds were aggregated to compare the candidate models in a statistically consistent manner. Model selection was based primarily on recall, with secondary consideration given to stability across folds and complementary metrics such as balanced accuracy and AUROC. This selection criterion reflects the prioritization-oriented formulation of the task, in which the goal is to identify candidates for further study within the heuristic labeling framework. Based on this procedure, a single classifier architecture and associated hyperparameter configuration were selected as the supervised baseline. This model was subsequently retrained on the full training set using the same preprocessing pipeline. The held-out test set was reserved exclusively for final evaluation and was not used during any stage of model selection or hyperparameter tuning.

To further assess the robustness and interpretability of the supervised baseline, three complementary analyses were performed: (i) feature ablation, (ii) perturbation-based sensitivity testing, and (iii) validation against an independent proxy habitability metric. In the feature ablation analysis, the full modeling pipeline---including scaling and classifier training---was repeated after excluding orbital eccentricity to evaluate the impact of its imputation and associated uncertainty on the classification task. Model performance was then compared with that obtained using the full feature set to determine the extent to which eccentricity contributes to the learned decision boundary. In the perturbation-based sensitivity analysis, a subset of input features---stellar mass, orbital period, and eccentricity---were subjected to small stochastic perturbations designed to approximate typical observational uncertainties. For a sample of planets ($\sim 100$), perturbed feature vectors were generated by sampling from Gaussian distributions centered on the nominal values, with standard deviations chosen to reflect measurement uncertainties. The errors were propagated to any features that were recalculated in the imputation stage (see Section~\ref{subsec:preprocessing}). The trained model was then evaluated on these perturbed datasets, and the variability of the resulting predictions and performance metrics was used to quantify sensitivity to input uncertainty.
	
Finally, to assess whether the model captures structure beyond the HWC labeling scheme, its predictions were computed against an independent proxy habitability metric constructed from physically motivated parameters and defined as
\begin{equation}
	H_{\text{proxy}} = \exp\left( -\sum_i \left( \frac{x_i - x_{i,\oplus}}{\sigma_i} \right)^2 \right),
\end{equation}
where the index $i$ runs over the set of planetary properties $\{R, M, F, T\}$, corresponding to planetary radius and mass, incident stellar flux, and equilibrium temperature. The reference values $x_{i,\oplus}$ represent Earth-like conditions ($R_\oplus = 1$, $M_\oplus = 1$, $F_\oplus = 1$, $T_\oplus = 255\,\mathrm{K}$), and the characteristic widths were set to $\sigma_R = 0.5$, $\sigma_M = 0.5$, $\sigma_F = 0.75$, and $\sigma_T = 50\,\mathrm{K}$. This formulation corresponds to a Gaussian-like similarity measure in a normalized parameter space and provides a smooth, continuous proxy for Earth-like conditions. It does not incorporate additional physical factors such as atmospheric composition, magnetic fields, or stellar activity, and therefore does not constitute a physically complete habitability model. Instead, it is used as a simplified, physically motivated reference for assessing consistency with the ML predictions. Correlation and overlap analyses were used to quantify the degree of agreement between the two approaches. These analyses are designed to evaluate the extent to which the model's behavior is robust to input uncertainties, sensitive to specific features, and consistent with independent physical proxies, while remaining explicitly conditioned on the heuristic labeling framework.

To facilitate physical interpretation of the selected baseline model in terms of the learned decision structure, two complementary feature importance analyses were performed. First, SHAP (SHapley Additive exPlanations) \citep{Lundberg2017} values were computed to quantify the average marginal contribution of each feature to the model's predictions. These contributions should be interpreted as reflecting the structure of the heuristic labeling scheme rather than direct physical drivers of habitability. SHAP values are grounded in cooperative game theory and provide consistent, model-agnostic explanations of feature influence. Second, permutation importance was computed with respect to recall. In this approach, the values of individual features are randomly permuted while all other features are held fixed, and the resulting change in recall is measured. This method directly assesses the dependence of model sensitivity on each feature. In particular, features such as ESI may encode composite heuristic information, and their importance reflects this constructed role within the labeling scheme. All the results of the analyses described above, alongside the corresponding visualizations are presented in Section~\ref{subsec:res-ml}.

The supervised learning baseline established in this section provides the fixed classifier architecture used throughout the AL experiments described in Section~\ref{subsec:AL}. By holding the model architecture constant, differences in performance can be attributed directly to the choice of query strategy rather than to changes in model capacity or optimization procedure. This design enables a direct assessment of the extent to which AL improves label efficiency within the constraints of the adopted heuristic classification framework.

\subsection{Active learning framework and query strategies}
\label{subsec:AL}
AL is employed in this study to reduce the labeling effort required to learn and reproduce the classification of potentially habitable exoplanets under the adopted labeling scheme by iteratively selecting the most informative candidates for annotation. We adopt a pool-based active learning \citep{Cohn1996} setting, which is appropriate when a large collection of unlabeled instances is available and labels are assumed to be costly, as is the case for habitability assessments that typically require detailed observational follow-up. In this context, labels correspond to the heuristic classifications described in Section~\ref{subsec:problem}, and the objective is to improve the efficiency of learning this mapping rather than to infer physical habitability directly.

The final preprocessed dataset described in Sections~\ref{subsec:data} and~\ref{subsec:preprocessing} is first divided into disjoint training and test subsets using an $80/20$ split, stratified by the habitability label to preserve the severe class imbalance. The test set remains fixed throughout all AL experiments and is never queried or augmented, serving exclusively as an unbiased evaluation set. The training subset is treated as the AL pool, from which labeled and unlabeled subsets are dynamically constructed during each run. At the start of every run, a small labeled seed is sampled from the training data, while the remaining instances constitute the unlabeled pool. This process is repeated independently across multiple runs to assess robustness against random initialization effects.

Seed selection is a critical design choice in AL, particularly under class imbalance. For each run, the initial labeled set consists of a total of 20 instances, including 3 planets labeled as potentially habitable and 17 labeled as non-habitable. This constraint ensures that the classifier is exposed to at least a minimal representation of the positive class from the outset, preventing degenerate early models that fail to identify rare positives. This design ensures that the model can begin to approximate the minority class defined by the labeling scheme from the earliest iterations. Seed instances are sampled randomly within each class from the training subset, using a different random seed for each run. This strategy balances realism---reflecting limited prior knowledge---with the practical necessity of initializing a usable classifier.

At the beginning of each run, feature scaling is performed using a robust scaler. The scaler is fitted exclusively on the seed set and subsequently applied to the unlabeled pool and the fixed test set, ensuring that no information from unlabeled or test instances leak into the scaling process. The classifier architecture used within the AL loop is fixed to the best-performing supervised model identified by the steps described in Section~\ref{subsec:ML} with tuned hyperparameters. At each iteration, the model is retrained from scratch on the expanded labeled set to reflect the updated training distribution.

Two query strategies are considered in this work: random sampling and margin sampling. Random sampling serves as a baseline, in which unlabeled instances are selected uniformly at random from the pool. Although simple, this strategy provides a reference against which the effectiveness of more informed querying can be evaluated. Margin sampling \citep{Balcan2007} is an uncertainty-based strategy that selects the instance for which the classifier is least confident. For probabilistic binary classifiers, this is quantified as the absolute difference between the predicted class probabilities. Instances with small margins lie closest  to the decision boundary and are therefore expected to be the most informative for refining the classifier. In this setting, informativeness is defined with respect to improving the classifier's approximation of the heuristic decision boundary. Margin sampling represents a purely exploitative strategy, focusing on regions of maximal uncertainty in the current model. Both strategies operate within the same AL loop and are evaluated under identical conditions to ensure a fair comparison.

For each run and each query strategy, the AL process proceeds iteratively for a fixed labeling budget of 70 queried instances. At every iteration, a single unlabeled instance is selected according to the AL strategy, removed from the pool, added to the labeled set along with its true label, and the classifier is retrained. Model performance is not evaluated after every query but instead recorded at regular intervals of five newly labeled instances. This aggregation interval reduces noise in the learning curves while preserving temporal resolution sufficient to capture performance trends. The stopping criterion is therefore budget-based rather than performance based, allowing learning trajectories to be compared across strategies. To account for stochastic variability arising from seed selection and model training, the entire procedure is repeated across 20 independent runs. Performance metrics are averaged across runs at each evaluation point, yielding mean learning curves and associated variability estimates.

At the conclusion of each run, the final trained model and its associated scaler are saved. These persisted models are subsequently reused for downstream analysis, including ensemble-based probability aggregation and planet recommendation, which are described in Section~\ref{subsec:recommend}. Importantly, this reuse does not affect the AL process itself, as all querying and evaluation decisions are completed prior to model persistence.

\subsection{Planet recommendation via ensemble uncertainty}
\label{subsec:recommend}
Beyond performance evaluation, an important objective of this study is to demonstrate how the trained models can be used to prioritize exoplanets for further observational follow-up. To this end, we formulate planet recommendation as a probabilistic ranking problem that explicitly accounts for model uncertainty arising from limited labeled data and stochastic training effects.

In the context of exoplanet habitability, false negatives are particularly costly, as potentially habitable planets may be overlooked due to incomplete or noisy measurements. Consequently, rather than relying on a single deterministic model prediction, we adopt an ensemble-based probabilistic approach that aggregates information across multiple independently trained models. Each AL run produces a final classifier trained on a distinct labeled set obtained through different random seeds and query trajectories. These models represent alternative hypotheses consistent with the data and the AL process. Their collective behavior provides a natural mechanism for quantifying uncertainty due to limited labeled data.

For each completed AL run, the final trained classifier and its associated feature scaler are retained. All retained models are subsequently applied to the entire dataset, including both the originally labeled instances and the remaining planets labeled as non-habitable in the catalog. Feature scaling is performed using the scaler associated with each individual model to preserve consistency with the model's training distribution. For a given planet, this procedure yields a set of predicted habitability probabilities, one from each model in the ensemble. No model refitting or additional calibration is performed at this stage.

The ensemble predictions are summarized using two statistics: the mean predicted probability and the standard deviation across models. The mean probability serves as an estimate of the model-averaged likelihood that a planet is classified as potentially habitable under the adopted labeling scheme, while the standard deviation provides a measure of uncertainty associated with this estimate. A high mean probability accompanied by low variability across models indicates a robust recommendation, whereas large variability suggests sensitivity to training data composition and reduced confidence. This distinction is particularly important when prioritizing targets for follow-up observations under constrained observational resources.

To focus the recommendation process on genuinely novel targets, only planets labeled as non-habitable in the original catalog are considered at this stage. These planets are ranked in descending order according to their ensemble-averaged habitability probability. No hard probability threshold is imposed a priori. Instead, the ranked list is inspected to identify planets that simultaneously exhibit elevated predicted habitability and relatively low ensemble uncertainty. This approach avoids overconfidence in marginal cases while retaining flexibility in defining follow-up priorities based on available resources or scientific objectives.

It is important to emphasize that the recommendation procedure does not assert definitive habitability. Rather, it provides a data-driven prioritization informed by the learned relationships in the available feature space. The resulting candidates should be interpreted as targets for further study, where additional observational constraints may confirm or refute their habitability status. Moreover, the recommendations are inherently conditioned on the choice of features, preprocessing steps, and labeling assumptions described in earlier sections. As such, they should be considered complementary to, rather than a replacement for, domain-driven assessment and physical modeling.

\section{Results}
\label{sec:res}
\subsection{Supervised baseline performance}
\label{subsec:res-ml}
\begin{figure}
	\centering
	\includegraphics[width=\columnwidth]{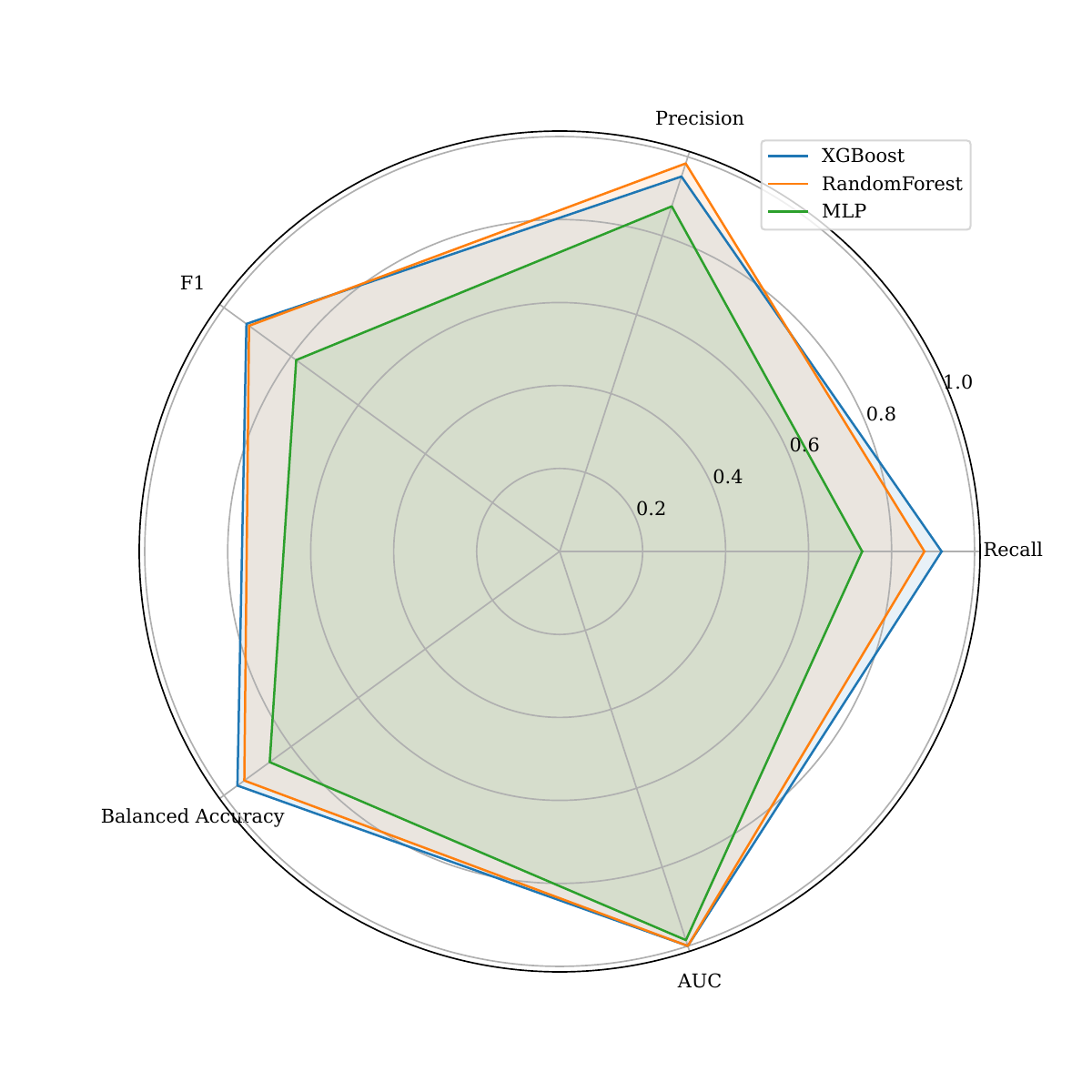}
	\caption{Radar plot comparing supervised classifiers across recall, precision, F1-score, balanced accuracy, and AUROC. XGBoost exhibits the strongest overall performance and is selected as the baseline model.}\label{fig:spider}
\end{figure}

We evaluated three supervised classifiers (RF, XGBoost, and MLP) using the metrics defined in Section~\ref{subsec:metrics}. Across repeated outer folds, XGBoost consistently achieved the highest mean recall, followed by RF, with the MLP showing both lower performance and substantially higher variance. To contextualize model performance beyond recall alone, we compare the three supervised classifiers across all evaluation metrics using a radar (spider) plot (Figure~\ref{fig:spider}). This visualization highlights the trade-offs between sensitivity, precision, and overall discrimination. While RF performs competitively on precision and balanced accuracy, XGBoost achieves the most favorable balance across all metrics simultaneously. The MLP underperforms consistently, particularly in recall and balanced accuracy, confirming its unsuitability as a baseline in this setting. Given these results, XGBoost is adopted as the reference supervised learner for all subsequent AL experiments.

The optimal configuration favored shallow trees and moderate subsampling, consistent with controlling overfitting in a low-signal, high-imbalance regime. The selected hyperparameters were a maximum tree depth of three, subsample and column-subsample fractions of 0.8, and a learning rate of 0.1. The model was then retrained on the full training set using these hyperparameters and evaluated once on a held-out test set that was not used during any stage of model selection. The quantitative performance of the final XGBoost model on the held-out test set is summarized in Table~\ref{tab:xgb}. It demonstrates high sensitivity to potentially habitable planets, while maintaining excellent discrimination overall, as reflected by the near-unity AUROC. These results establish XGBoost as a strong supervised reference and justify its use as the base learner in the AL framework described in Section~\ref{subsec:AL}. It is noteworthy that recall, precision, and F1-score take identical numerical values for the final XGBoost model. This occurs because, at the selected operating point on the held-out test set, the number of false positives and false negatives is equal, yielding identical values for precision and recall. Since the F1-score is defined as the harmonic mean of precision and recall, it reduces to the same value under these conditions. This equality is therefore a consequence of the specific confusion-matrix configuration induced by the decision threshold, rather than an artifact of the evaluation process.

\begin{table}
	\centering
	\caption{Performance of the final XGBoost classifier evaluated on the held-out test set. Metrics are defined in Section~\ref{subsec:metrics}.}\label{tab:xgb}
	\begin{tabular}{lc}
		\hline\hline
		\textbf{Metric} 	& \textbf{Value}	\\
		\hline
		Recall 				& 0.929				\\
		Precision 			& 0.929				\\
		F1-score 			& 0.929				\\
		Balanced accuracy 	& 0.964				\\
		AUROC 				& 0.999				\\
		\hline
	\end{tabular}
\end{table}

\begin{figure}
	\centering
	\includegraphics[width=\columnwidth]{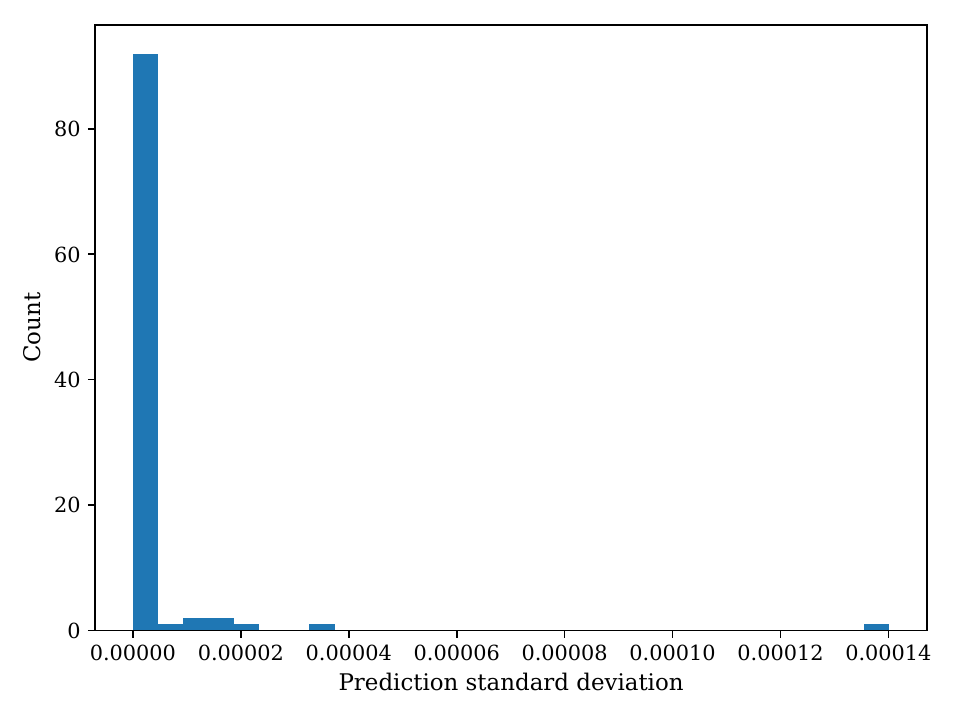}
	\caption{Distribution of prediction variability under input perturbations, quantified as the standard deviation of model-predicted habitability probabilities across perturbed realizations.}\label{fig:perturbation}
\end{figure}

To assess the impact of eccentricity and its imputation on the classification, we compared the final XGBoost model with and without this feature. The resulting performance remained unchanged to three significant figures, indicating that its contribution is negligible within the adopted feature set. The implications of this result are discussed in Section~\ref{sec:disc}. We further examined the sensitivity of the model to input uncertainties by evaluating the variability of predicted probabilities under realistic perturbations of the input features. The resulting prediction variability is extremely small, with a median standard deviation of $1.46 \times 10^{-7}$ and a maximum of $1.24 \times 10^{-4}$. The distribution of these values (Figure~\ref{fig:perturbation}) shows that the vast majority of predictions are effectively invariant to perturbations. To evaluate whether the model captures structure beyond the heuristic labeling scheme, we compared its predicted probabilities with an independent proxy habitability metric. A moderate positive correlation is observed (Pearson $r = 0.383$, Spearman $\rho = 0.292$), with substantial scatter.

\begin{figure}
	\centering
	\includegraphics[width=\columnwidth]{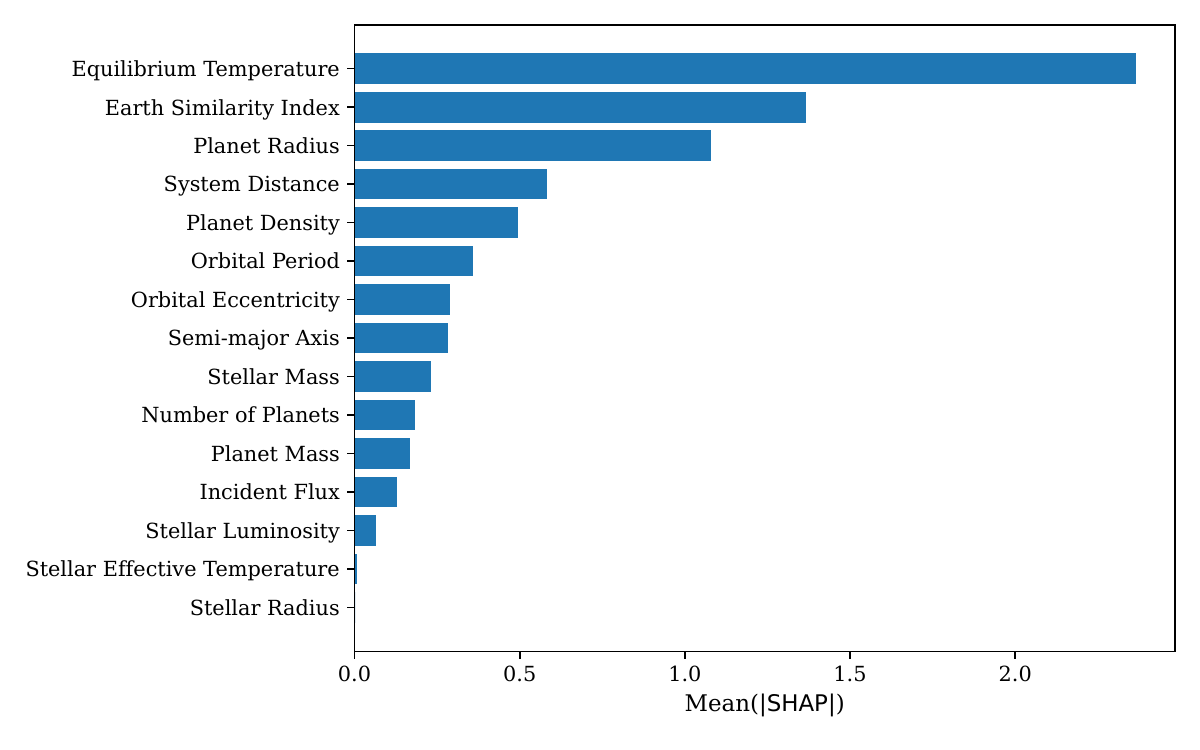}
	\caption{Global feature importance based on mean absolute SHAP values for the XGBoost classifier. Features are ordered by decreasing contribution to the model output, highlighting the dominant role of thermal and habitability-related parameters.}\label{fig:shap}
\end{figure}

\begin{figure}
	\centering
	\includegraphics[width=\columnwidth]{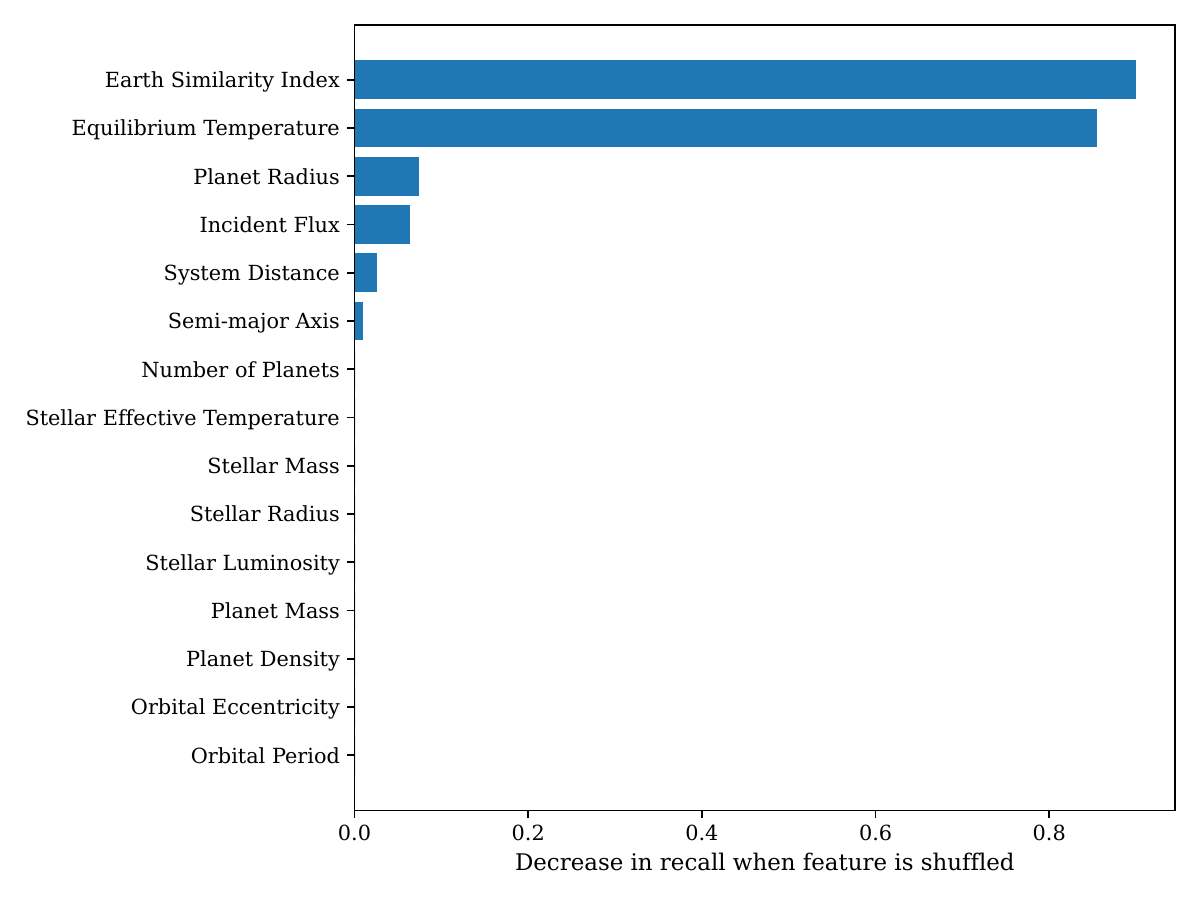}
	\caption{Permutation feature importance measured as the decrease in recall when individual features are randomly shuffled. Larger values indicate greater importance for identifying potentially habitable planets.}\label{fig:perm}
\end{figure}

To interpret the predictions of the selected XGBoost model, we employed two complementary feature importance techniques. First, SHAP values were used to quantify the average marginal contribution of each feature to the model output. The global SHAP summary (Figure~\ref{fig:shap}) indicates that planetary equilibrium temperature and ESI dominate the model's decision process, followed by planetary radius and system distance. Orbital parameters and stellar properties contribute more weakly but non-negligibly. Second, permutation importance was computed with respect to recall by randomly shuffling individual features and measuring the resulting decrease in recall. The permutation results (Figure~\ref{fig:perm}) reinforce the SHAP findings, highlighting equilibrium temperature and ESI as the most critical predictors for recall, with a steep performance drop when either feature is disrupted.

\subsection{Active learning performance}
\label{subsec:res-al}
\begin{figure*}
	\centering
	\includegraphics[width=\textwidth]{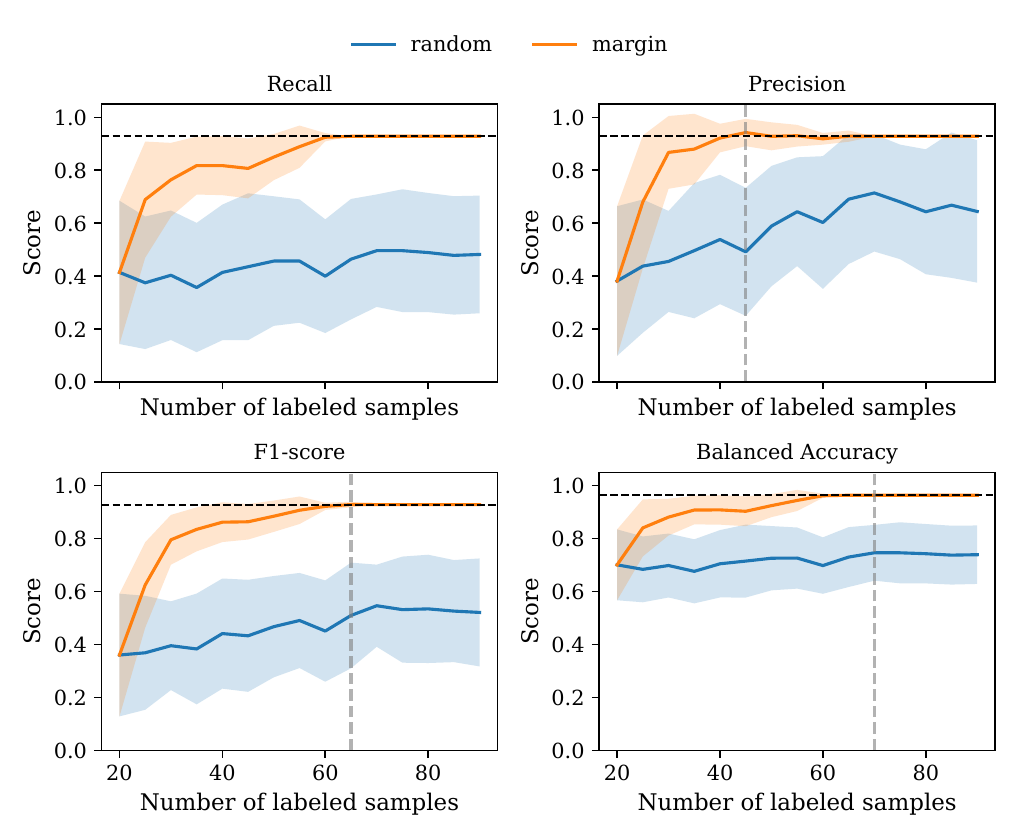}
	\caption{Active learning performance as a function of the number of labeled instances. Mean learning curves over 20 independent runs, evaluated every five queried instances. Shaded regions indicate $\pm1$ standard deviation across runs, illustrating inter-run variability. The horizontal dashed lines indicate the performance of the XGBoost classifier trained on the full labeled training pool. The vertical dashed lines mark the number of labeled instances at which the margin-based strategy approaches this performance level for each metric.}\label{fig:performance}
\end{figure*}

To evaluate the effectiveness of AL relative to passive data acquisition, we compared a random sampling baseline with an uncertainty-driven margin sampling strategy. The learning dynamics are summarized in Figure~\ref{fig:performance}, which shows the evolution of recall, precision, F1-score, and balanced accuracy as a function of the number of labeled instances. Shaded regions indicate the inter-run variability, quantified as one standard deviation around the mean. A horizontal reference level is included to indicate the performance of the XGBoost classifier trained on the full labeled training pool.

From the earliest stages of the learning process, margin-based querying substantially outperforms random sampling in terms of recall. With only 20 labeled instances (the initial seed set), both strategies start from comparable performance levels. However, after a small number of AL queries, the margin strategy rapidly identifies informative samples near the classifier's decision boundary, leading to a sharp increase in recall. The early-recall statistic, computed after the first evaluation interval, reaches 0.925 for margin sampling compared to 0.459 for random sampling, demonstrating a more than twofold improvement in the recovery of potentially habitable planets under severe label scarcity. As additional instances are acquired, the superiority of margin sampling persists across all reported metrics. Precision and F1-score increase steadily. By the later stages of the budget, the margin strategy converges to high and stable performance, with late-stage precision and F1-scores both reaching approximately 0.929. In contrast, random sampling exhibits slower improvement and substantially larger variance across runs.

Balanced accuracy, which accounts for both sensitivity and specificity under class imbalance, further highlights the benefits of active querying. The margin strategy attains a late-stage balanced accuracy of 0.964, whereas random sampling remains below 0.741 even after exhausting the labeling budget. An important qualitative feature of the learning curves is the early saturation of the margin strategy. After approximately 60--65 labeled instances, performance gains plateau across all metrics. The margin sampler reaches high recall and balanced accuracy while approaching the performance of the XGBoost classifier trained on the full labeled training pool using only $\sim$60--65 labeled instances. In relative terms, this corresponds to achieving near-supervised performance using only a few percents of the labeled training pool. While near-convergence to the full data performance level is achieved for precision, F1-score, and balanced accuracy, recall remains slightly below the corresponding threshold.
\subsection{Planet recommendation}
\label{subsec:res-rec}
To translate the outcomes of the AL framework into an astronomically meaningful product, we applied the ensemble of trained classifiers to the full preprocessed dataset to identify candidate planets for follow-up study based on their predicted habitability probabilities. The ranked list of candidates revealed a single planet that clearly separated from the rest of the non-habitable population. This exoplanet exhibited a mean predicted habitability probability of 0.82 with a standard deviation of 0.06 across the ensemble, while remaining candidates showed substantially lower mean probabilities ($\le0.30$).

Cross-matching this candidate against the merged dataset of the HWC and the PSCompPars table using tolerance-based matching on orbital, planetary, and stellar parameters yielded a unique identification: $\tau$ Ceti f, a planet orbiting the nearby G-type \citep{Turnbull2015} star $\tau$ Ceti at a distance of 3.6 pc \citep{Stassun2019}. $\tau$ Ceti f was discovered via the radial-velocity method in 2017 and resides in a multi-planet system with four known planets \citep{Feng2017}. The physical parameters of $\tau$ Ceti f place it within the broader distribution of the potentially habitable sample, while deviating from its central tendency in key parameters such as incident flux and equilibrium temperature. Its planetary radius ($1.81 R_\oplus$) and mass ($3.93 M_\oplus$) are consistent with the upper range of super-Earths, while its orbital semi-major axis (1.33 AU) and equilibrium temperature ($\approx 185$ K) lie within the overall distribution of cataloged potentially habitable planets but deviate from their central tendency. A comparison with the median and interquartile range of the potentially habitable population is provided in Table~\ref{tab:comparison}. Notably, the relatively low incident flux and equilibrium temperature place $\tau$ Ceti f outside the conventional habitable zone for liquid surface water under standard assumptions, highlighting a tension between the model-based ranking and simplified physical criteria.

\setlength{\tabcolsep}{2pt}
\begin{table}
	\centering
	\caption{Comparison between the recommended planet ($\tau$ Ceti f) and the distribution of cataloged potentially habitable planets. For each parameter, the value of the recommended candidate is shown alongside the median and interquartile range (IQR) of the potentially habitable sample.}
	\label{tab:comparison}
	\begin{tabular}{lccc}
		\hline
		\hline
		\multicolumn{4}{c}{\textit{Planetary and orbital parameters}} \\
		\hline
		\textbf{Parameter} & \textbf{Candidate} & \textbf{Median} & \textbf{IQR} \\
		\hline
		Orbital period (days)              & 636.13 & 36.77 & 92.30 \\
		Semi-major axis (AU)               & 1.334  & 0.165 & 0.290 \\
		Orbital eccentricity               & 0.160  & 0.083 & 0.141 \\
		Planet radius ($R_\oplus$)         & 1.81   & 1.78  & 0.90  \\
		Planet mass ($M_\oplus$)           & 3.93   & 4.06  & 3.48  \\
		Planet density (g cm$^{-3}$)       & 0.66   & 0.68  & 0.46  \\
		Incident flux ($S_\oplus$)         & 0.28   & 0.85  & 0.68  \\
		Equilibrium temperature (K)        & 184.7  & 244.0 & 47.4  \\
		Earth Similarity Index             & 0.555  & 0.729 & 0.169 \\
		\hline
		\multicolumn{4}{c}{\textit{Host star and system parameters}} \\
		\hline
		Stellar $T_{\mathrm{eff}}$ (K)     & 5310   & 3460  & 919.5 \\
		Stellar mass ($M_\odot$)           & 0.78   & 0.42  & 0.36  \\
		Stellar radius ($R_\odot$)         & 0.83   & 0.42  & 0.34  \\
		Stellar luminosity ($L_\odot$)     & 0.50   & 0.023 & 0.072 \\
		System distance (pc)               & 3.60   & 31.13 & 241.09 \\
		Planets in system                  & 4      & 2     & 2     \\
		\hline
		\hline
	\end{tabular}
\end{table}

Inspection of instance-level SHAP values indicates that the elevated model score assigned to $\tau$ Ceti f is primarily associated with its incident stellar flux, planetary radius, and equilibrium temperature, with secondary contributions from orbital eccentricity and system distance. These factors collectively shift the model output toward the positive class despite the planet's exclusion from the catalog's potentially habitable designation.

Importantly, this recommendation does not constitute a reclassification of $\tau$ Ceti f as habitable. Instead, it reflects the model's learned representation of the heuristic labeling scheme within the available feature space. The apparent discrepancy between the model-based ranking and simplified physical expectations emphasizes the importance of interpreting such outputs as prioritization signals rather than physical determinations of habitability. In this sense, $\tau$ Ceti f serves as a concrete example of how an ensemble-based AL framework can guide follow-up investigations while remaining complementary to domain-driven assessment.

\section{Discussion}
\label{sec:disc}
This study investigated the use of AL for exoplanet habitability classification under conditions of extreme class imbalance and label uncertainty. By combining a carefully constructed feature set, a strong supervised baseline, and uncertainty-driven AL strategies, we evaluated both predictive performance and label efficiency, and demonstrated a concrete downstream application in the form of planet recommendation. The results should be interpreted primarily as a methodological assessment of how AL interacts with heuristic habitability labels, rather than as a direct probe of the underlying physics of planetary habitability. Below, we discuss the implications of these findings, their relevance to habitability studies, and the limitations that contextualize the conclusions.

The results presented in Section~\ref{subsec:res-al} demonstrate that margin-based AL substantially outperforms random sampling throughout the labeling process, particularly in the low-label regime most relevant to astronomical follow-up. The early-recall improvement observed for margin sampling highlights a central strength of uncertainty-based AL: the ability to focus labeling effort on ambiguous regions of feature space where the classifier's decision boundary is poorly constrained. Under extreme class imbalance, such regions are disproportionately informative for identifying rare positive instances. An important qualitative feature of the learning curves is the rapid saturation of margin sampling relative to random sampling. Performance gains plateau after approximately 60--65 labeled instances, suggesting that most of the discriminative structure present in the adopted feature space can be captured efficiently when queries are selected adaptively. This behavior is consistent with the concentration of informative instances near the class boundary, where uncertainty-based querying is most efficient. In contrast, random sampling does not achieve comparable performance within the same labeling budget, reflecting its inefficiency in discovering informative minority-class examples when positives are sparse. Notably, margin sampling approaches---but does not exceed---the performance of the fully supervised XGBoost baseline trained on the entire training set. This behavior is expected: AL does not introduce new information beyond what is present in the data, but instead reorders label acquisition. The convergence of AL performance toward the supervised upper bound therefore supports the interpretation that AL achieves label efficiency rather than absolute performance gains, reaching near-supervised performance with substantially fewer labeled instances. The slight underperformance in recall relative to the full-data baseline suggests that complete recovery of the minority class remains sensitive  to additional labeled examples, reflecting the intrinsic difficulty of identifying rare potentially habitable planets under strong class imbalance.

From an astrophysical perspective, the results primarily reflect the structure of the underlying catalog definitions rather than independent physical inference. The substantial overlap between potentially habitable and non-habitable planets in marginal feature distributions (Figure~\ref{fig:features}) explains why uncertainty-driven AL is particularly effective: planets near the decision boundary are both common and informative, and querying them accelerates learning. The dominance of equilibrium temperature, incident stellar flux, and ESI in both SHAP and permutation importance analyses suggests that the models largely recover relationships already encoded---implicitly or explicitly---in the HWC labeling scheme. However, the presence of secondary contributions from orbital eccentricity, system distance, and planetary bulk properties indicates that the classifier integrates multiple weak signals rather than relying on a single proxy. An additional insight arises from the negligible impact of orbital eccentricity on classification performance (Section~\ref{subsec:res-ml}). Although eccentricity is physically relevant to long-term climate variability and habitability, its exclusion from the feature set does not measurably degrade model performance. This behavior suggests that, within the adopted feature set, the information carried by eccentricity is largely redundant with flux- and temperature-based parameters, which already encode the dominant effects of orbital geometry on planetary irradiation. Importantly, this result should not be interpreted as indicating that eccentricity is physically unimportant, but rather that its predictive contribution is subsumed by other features under the present heuristic labeling scheme and data constraints. This behavior is desirable in a prioritization context, where incomplete, uncertain, or partially redundant measurements are the norm, and robust performance depends on the model's ability to extract information from the most informative features while remaining insensitive to less predictive or noisier parameters. 
In addition, tree-based ensemble methods outperform neural networks in this setting, likely due to their robustness to heterogeneous feature scales and limited positive samples.

Within this context, the role of AL is not to redefine habitability, but to optimize how information is extracted from an existing labeling framework. Importantly, it provides a principled mechanism for prioritizing planets that are ambiguous under current catalog definitions. In this sense, AL aligns naturally with observational practice: it identifies targets where additional data are most likely to refine scientific understanding, rather than reinforcing already well-separated cases. To further assess the extent to which the model captures information beyond the adopted labeling scheme, we compared its predictions to an independent proxy habitability metric. The resulting moderate correlation, accompanied by substantial scatter, indicates partial alignment with physically motivated trends while demonstrating that the model does not simply reproduce the proxy definition.

The ensemble-based recommendation of $\tau$ Ceti f illustrates the conservative character of the proposed framework. Despite evaluating thousands of non-habitable-labeled planets, only a single candidate emerges with both a high mean predicted habitability probability and low-model uncertainty. This outcome reflects the combined effects of severe class imbalance, cautious aggregation across independently trained models, and the requirement for consistency rather than isolated high scores. Such conservatism is a feature rather than a limitation in the context of follow-up prioritization. False positives in habitability studies can divert scarce observational resources, whereas false negatives can often be revisited as catalogs evolve. By emphasizing robustness across models and explicitly quantifying uncertainty, the recommendation procedure favors candidates whose prioritization is stable under variations in training data and AL trajectories. The identification of $\tau$ Ceti f---already a well-studied planet in a nearby system \citep{Feng2017}---also highlights an important interpretive point: the framework is not designed to discover exotic outliers, but rather to reassess borderline cases in light of learned multivariate patterns. However, the candidate's physical properties, particularly its relatively low incident and equilibrium temperature, place it outside the conventional habitable zone under standard assumptions. This discrepancy emphasizes that the model-based ranking reflects consistency with the learned heuristic decision space rather than a direct indication of physical habitability.

Several limitations should be borne in mind when interpreting the results. First, the habitability labels used throughout the study are heuristic and literature-derived, rather than direct indicators of biological or even surface habitability. While AL is robust to some degree of label noise, systematic biases in catalog definitions inevitably propagate into the learned models. In particular, the reliance on composite indices such as the ESI introduces implicit correlations that the model may reproduce rather than independently infer. Second, the feature set, while physically motivated, is constrained by current data availability. Atmospheric composition, obliquity, magnetic fields, and long-term climate stability are not represented, yet are critical to habitability. The conclusions therefore apply strictly within the scope of the adopted features and should not be extrapolated beyond them. Third, input uncertainties are not explicitly propagated through the model. Parameters such as semi-major axis and eccentricity can carry substantial observational uncertainties, and the reported classifications should therefore be interpreted as conditional on nominal values. While we performed a sampling-based perturbation analysis to assess the sensitivity of model predictions to input uncertainties, this approach is limited to local perturbations around nominal values. Extending this analysis to a more systematic uncertainty propagation framework, for example through Monte Carlo sampling of full parameter distributions, would provide a more comprehensive assessment of how observational uncertainties impact classification outcomes. Fourth, the AL simulations assume oracle access to true labels during querying. In practice, follow-up observations yield partial, delayed, or uncertain information rather than binary labels. Extending the framework to incorporate probabilistic or multi-fidelity labels represents an important direction for future work. Finally, the recommendation results are conditioned on the specific classifier architecture (XGBoost) and query strategy (margin sampling) selected in this study. While the consistency of results across runs suggests robustness, alternative models or strategies could yield different prioritization, particularly as new data become available.

\section{Conclusions}
\label{sec:conc}
In this work, we investigated the use of AL to accelerate habitability classification in exoplanet catalogs characterized by extreme class imbalance, heterogeneous data quality, and uncertain labels. By combining physically motivated feature selection, careful preprocessing, and a robust supervised baseline with uncertainty-driven query strategies, we demonstrated that AL can substantially reduce the number of labeled instances required to achieve near-supervised performance.

Beyond performance evaluation, we demonstrated a practical downstream application of the AL framework by aggregating predictions from multiple trained models to recommend candidates for further observational study. By ranking non-habitable-labeled planets according to ensemble mean probability and associated uncertainty, we identified a single robust candidate---$\tau$ Ceti f---with a high predicted habitability probability and low inter-model variance. This result illustrates how AL outputs can be translated into conservative, uncertainty-aware prioritization. Importantly, such recommendations reflect consistency within the learned decision space and should be interpreted in conjunction with physical constraints rather than as standalone indicators of habitability.

Overall, this study shows that AL provides a principled and computationally tractable approach for guiding follow-up efforts in exoplanet science, particularly as catalogs continue to grow faster than observational resources. While the conclusions are necessarily conditioned on heuristic habitability labels and currently available features, the proposed framework is readily extensible to future datasets, alternative classifiers, and more nuanced labeling schemes. Future work incorporating explicit uncertainty propagation and validation against more physically grounded habitability models will be essential for strengthening the connection between data-driven classification and astrophysical interpretation. As exoplanet surveys expand in scale and complexity, integrating AL into catalog analysis pipelines offers a promising path toward more efficient and transparent scientific decision-making.


\begin{acknowledgments}
This research has made use of the NASA Exoplanet Archive, which is operated by the California Institute of Technology, under contract with the National Aeronautics and Space Administration under the Exoplanet Exploration Program.

We are grateful to the anonymous referee for their insightful comments and suggestions, which have significantly improved the quality and clarity of this manuscript.
\end{acknowledgments}

%
%
%

%
\facility{Exoplanet Archive}

\software{numpy \citep{Walt2011}, 
          scipy \citep{Virtanen2020},
          scikit-learn \citep{Pedregosa2011},
          xgboost \citep{Chen2016},
          shap \citep{Lundberg2017},
          matplotlib \citep{Hunter2007},
          upsetplot \citep{Nothman2018},
          seaborn \citep{Waskom2021},
          modAL \citep{Danka2018}
          }

\bibliography{refs}{}

@Misc{HWC2025,
  author       = {{PHL @ UPR Arecibo}},
  howpublished = {\url{http://phl.upr.edu/hwc}},
  month        = {October},
  title        = {The Habitable Worlds Catalog (HWC)},
  year         = {2025},
  day          = {22},
}

@Article{Christiansen2025,
  author    = {Christiansen, Jessie L. and McElroy, Douglas L. and Harbut, Marcy and Ciardi, David R. and Crane, Megan and Good, John and Hardegree-Ullman, Kevin K. and Kesseli, Aurora Y. and Lund, Michael B. and Lynn, Meca and Muthiar, Ananda and Nilsson, Ricky and Oluyide, Toba and Papin, Michael and Rivera, Amalia and Swain, Melanie and Susemiehl, Nicholas D. and Tam, Raymond and van Eyken, Julian and Beichman, Charles},
  journal   = {Planet. Sci. J.},
  title     = {The NASA Exoplanet Archive and Exoplanet Follow-up Observing Program: Data, Tools, and Usage},
  year      = {2025},
  issn      = {2632-3338},
  month     = aug,
  number    = {8},
  pages     = {186},
  volume    = {6},
  doi       = {10.3847/psj/ade3c2},
  publisher = {American Astronomical Society},
}

@Article{Walt2011,
  author    = {van der Walt, Stefan and Colbert, S. Chris and Varoquaux, Gael},
  journal   = {Comput. Sci. Eng.},
  title     = {The NumPy Array: A Structure for Efficient Numerical Computation},
  year      = {2011},
  issn      = {1558-366X},
  month     = mar,
  number    = {2},
  pages     = {22--30},
  volume    = {13},
  doi       = {10.1109/mcse.2011.37},
  publisher = {Institute of Electrical and Electronics Engineers (IEEE)},
}

@Article{Virtanen2020,
  author    = {Virtanen, Pauli and Gommers, Ralf and Oliphant, Travis E. and Haberland, Matt and Reddy, Tyler and Cournapeau, David and Burovski, Evgeni and Peterson, Pearu and Weckesser, Warren and Bright, Jonathan and van der Walt, Stéfan J. and Brett, Matthew and Wilson, Joshua and Millman, K. Jarrod and Mayorov, Nikolay and Nelson, Andrew R. J. and Jones, Eric and Kern, Robert and Larson, Eric and Carey, C J and Polat, İlhan and Feng, Yu and Moore, Eric W. and VanderPlas, Jake and Laxalde, Denis and Perktold, Josef and Cimrman, Robert and Henriksen, Ian and Quintero, E. A. and Harris, Charles R. and Archibald, Anne M. and Ribeiro, Antônio H. and Pedregosa, Fabian and van Mulbregt, Paul and Vijaykumar, Aditya and Bardelli, Alessandro Pietro and Rothberg, Alex and Hilboll, Andreas and Kloeckner, Andreas and Scopatz, Anthony and Lee, Antony and Rokem, Ariel and Woods, C. Nathan and Fulton, Chad and Masson, Charles and Häggström, Christian and Fitzgerald, Clark and Nicholson, David A. and Hagen, David R. and Pasechnik, Dmitrii V. and Olivetti, Emanuele and Martin, Eric and Wieser, Eric and Silva, Fabrice and Lenders, Felix and Wilhelm, Florian and Young, G. and Price, Gavin A. and Ingold, Gert-Ludwig and Allen, Gregory E. and Lee, Gregory R. and Audren, Hervé and Probst, Irvin and Dietrich, Jörg P. and Silterra, Jacob and Webber, James T and Slavič, Janko and Nothman, Joel and Buchner, Johannes and Kulick, Johannes and Schönberger, Johannes L. and de Miranda Cardoso, José Vinícius and Reimer, Joscha and Harrington, Joseph and Rodríguez, Juan Luis Cano and Nunez-Iglesias, Juan and Kuczynski, Justin and Tritz, Kevin and Thoma, Martin and Newville, Matthew and Kümmerer, Matthias and Bolingbroke, Maximilian and Tartre, Michael and Pak, Mikhail and Smith, Nathaniel J. and Nowaczyk, Nikolai and Shebanov, Nikolay and Pavlyk, Oleksandr and Brodtkorb, Per A. and Lee, Perry and McGibbon, Robert T. and Feldbauer, Roman and Lewis, Sam and Tygier, Sam and Sievert, Scott and Vigna, Sebastiano and Peterson, Stefan and More, Surhud and Pudlik, Tadeusz and Oshima, Takuya and Pingel, Thomas J. and Robitaille, Thomas P. and Spura, Thomas and Jones, Thouis R. and Cera, Tim and Leslie, Tim and Zito, Tiziano and Krauss, Tom and Upadhyay, Utkarsh and Halchenko, Yaroslav O. and Vázquez-Baeza, Yoshiki},
  journal   = {Nat. Methods},
  title     = {SciPy 1.0: fundamental algorithms for scientific computing in Python},
  year      = {2020},
  issn      = {1548-7105},
  month     = feb,
  number    = {3},
  pages     = {261--272},
  volume    = {17},
  doi       = {10.1038/s41592-019-0686-2},
  publisher = {Springer Science and Business Media LLC},
}

@Article{Pedregosa2011,
  author    = {Pedregosa, Fabian and Varoquaux, Ga\"{e}l and Gramfort, Alexandre and Michel, Vincent and Thirion, Bertrand and Grisel, Olivier and Blondel, Mathieu and Prettenhofer, Peter and Weiss, Ron and Dubourg, Vincent and Vanderplas, Jake and Passos, Alexandre and Cournapeau, David and Brucher, Matthieu and Perrot, Matthieu and Duchesnay, \'{E}douard},
  journal   = {J. Mach. Learn. Res.},
  title     = {Scikit-learn: Machine Learning in Python},
  year      = {2011},
  issn      = {1532-4435},
  pages     = {2825–2830},
  volume    = {12},
  publisher = {JMLR.org},
}

@InProceedings{Chen2016,
  author     = {Chen, Tianqi and Guestrin, Carlos},
  booktitle  = {Proceedings of the 22nd ACM SIGKDD International Conference on Knowledge Discovery and Data Mining},
  title      = {XGBoost: A Scalable Tree Boosting System},
  year       = {2016},
  month      = aug,
  pages      = {785--794},
  publisher  = {ACM},
  series     = {KDD ’16},
  collection = {KDD ’16},
  doi        = {10.1145/2939672.2939785},
}

@Article{Lundberg2017,
  author        = {Lundberg, Scott and Lee, Su-In},
  title         = {A Unified Approach to Interpreting Model Predictions},
  year          = {2017},
  month         = may,
  archiveprefix = {arXiv},
  copyright     = {arXiv.org perpetual, non-exclusive license},
  doi           = {10.48550/ARXIV.1705.07874},
  eprint        = {1705.07874},
  primaryclass  = {cs.AI},
  publisher     = {arXiv},
}

@Article{Hunter2007,
  author    = {Hunter, John D.},
  journal   = {Comput. Sci. Eng.},
  title     = {Matplotlib: A 2D Graphics Environment},
  year      = {2007},
  issn      = {1521-9615},
  number    = {3},
  pages     = {90--95},
  volume    = {9},
  doi       = {10.1109/mcse.2007.55},
  publisher = {Institute of Electrical and Electronics Engineers (IEEE)},
}

@Article{Waskom2021,
  author    = {Waskom, Michael},
  journal   = {JOSS},
  title     = {seaborn: statistical data visualization},
  year      = {2021},
  issn      = {2475-9066},
  month     = apr,
  number    = {60},
  pages     = {3021},
  volume    = {6},
  doi       = {10.21105/joss.03021},
  publisher = {The Open Journal},
}

@Article{Danka2018,
  author        = {Danka, Tivadar and Horvath, Peter},
  title         = {modAL: A modular active learning framework for Python},
  year          = {2018},
  month         = may,
  archiveprefix = {arXiv},
  copyright     = {arXiv.org perpetual, non-exclusive license},
  doi           = {10.48550/ARXIV.1805.00979},
  eprint        = {1805.00979},
  primaryclass  = {cs.LG},
  publisher     = {arXiv},
}

@Article{Richards2011,
  author    = {Richards, Joseph W. and Starr, Dan L. and Brink, Henrik and Miller, Adam A. and Bloom, Joshua S. and Butler, Nathaniel R. and Berian James, J. and Long, James P. and Rice, John},
  journal   = {ApJ},
  title     = {ACTIVE LEARNING TO OVERCOME SAMPLE SELECTION BIAS: APPLICATION TO PHOTOMETRIC VARIABLE STAR CLASSIFICATION},
  year      = {2011},
  issn      = {1538-4357},
  month     = dec,
  number    = {2},
  pages     = {192},
  volume    = {744},
  doi       = {10.1088/0004-637x/744/2/192},
  publisher = {American Astronomical Society},
}

@Article{Ishida2018,
  author    = {Ishida, E E O and Beck, R and González-Gaitán, S and de Souza, R S and Krone-Martins, A and Barrett, J W and Kennamer, N and Vilalta, R and Burgess, J M and Quint, B and Vitorelli, A Z and Mahabal, A and Gangler, E},
  journal   = {MNRAS},
  title     = {Optimizing spectroscopic follow-up strategies for supernova photometric classification with active learning},
  year      = {2018},
  issn      = {1365-2966},
  month     = nov,
  number    = {1},
  pages     = {2--18},
  volume    = {483},
  doi       = {10.1093/mnras/sty3015},
  publisher = {Oxford University Press (OUP)},
}

@Article{Skoda2020,
  author    = {Škoda, P. and Podsztavek, O. and Tvrdík, P.},
  journal   = {A\&A},
  title     = {Active deep learning method for the discovery of objects of interest in large spectroscopic surveys},
  year      = {2020},
  issn      = {1432-0746},
  month     = nov,
  pages     = {A122},
  volume    = {643},
  doi       = {10.1051/0004-6361/201936090},
  publisher = {EDP Sciences},
}

@Article{Solorio2005,
  author    = {Solorio, T. and Fuentes, O. and Terlevich, R. and Terlevich, E.},
  journal   = {MNRAS},
  title     = {An active instance-based machine learning method for stellar population studies},
  year      = {2005},
  issn      = {1365-2966},
  month     = oct,
  number    = {2},
  pages     = {543--554},
  volume    = {363},
  doi       = {10.1111/j.1365-2966.2005.09456.x},
  publisher = {Oxford University Press (OUP)},
}

@Article{Masters2015,
  author    = {Masters, Daniel and Capak, Peter and Stern, Daniel and Ilbert, Olivier and Salvato, Mara and Schmidt, Samuel and Longo, Giuseppe and Rhodes, Jason and Paltani, Stephane and Mobasher, Bahram and Hoekstra, Henk and Hildebrandt, Hendrik and Coupon, Jean and Steinhardt, Charles and Speagle, Josh and Faisst, Andreas and Kalinich, Adam and Brodwin, Mark and Brescia, Massimo and Cavuoti, Stefano},
  journal   = {ApJ},
  title     = {MAPPING THE GALAXY COLOR–REDSHIFT RELATION: OPTIMAL PHOTOMETRIC REDSHIFT CALIBRATION STRATEGIES FOR COSMOLOGY SURVEYS},
  year      = {2015},
  issn      = {1538-4357},
  month     = oct,
  number    = {1},
  pages     = {53},
  volume    = {813},
  doi       = {10.1088/0004-637x/813/1/53},
  publisher = {American Astronomical Society},
}

@Article{Hoyle2016,
  author    = {Hoyle, Ben and Paech, Kerstin and Rau, Markus Michael and Seitz, Stella and Weller, Jochen},
  journal   = {MNRAS},
  title     = {Tuning target selection algorithms to improve galaxy redshift estimates},
  year      = {2016},
  issn      = {1365-2966},
  month     = mar,
  number    = {4},
  pages     = {4498--4511},
  volume    = {458},
  doi       = {10.1093/mnras/stw563},
  publisher = {Oxford University Press (OUP)},
}

@Article{Ishida2021,
  author    = {Ishida, E. E. O. and Kornilov, M. V. and Malanchev, K. L. and Pruzhinskaya, M. V. and Volnova, A. A. and Korolev, V. S. and Mondon, F. and Sreejith, S. and Malancheva, A. A. and Das, S.},
  journal   = {A\&A},
  title     = {Active anomaly detection for time-domain discoveries},
  year      = {2021},
  issn      = {1432-0746},
  month     = jun,
  pages     = {A195},
  volume    = {650},
  doi       = {10.1051/0004-6361/202037709},
  publisher = {EDP Sciences},
}

@InProceedings{Kennamer2020,
  author    = {Kennamer, Noble and Ishida, Emille E. O. and Gonzalez-Gaitan, Santiago and Souza, Rafael S. de and Ihler, Alexander and Ponder, Kara and Vilalta, Ricardo and Moller, Anais and Jones, David O. and Dai, Mi and Krone-Martins, Alberto and Quint, Bruno and Sreejith, Sreevarsha and Malz, Alex I. and Galbany, Lluis},
  booktitle = {2020 IEEE Symposium Series on Computational Intelligence (SSCI)},
  title     = {Active learning with RESSPECT: Resource allocation for extragalactic astronomical transients},
  year      = {2020},
  month     = dec,
  pages     = {3115--3124},
  publisher = {IEEE},
  doi       = {10.1109/ssci47803.2020.9308300},
}

@Article{Leoni2022,
  author    = {Leoni, M. and Ishida, E. E. O. and Peloton, J. and Möller, A.},
  journal   = {A\&A},
  title     = {Fink: Early supernovae Ia classification using active learning},
  year      = {2022},
  issn      = {1432-0746},
  month     = jul,
  pages     = {A13},
  volume    = {663},
  doi       = {10.1051/0004-6361/202142715},
  publisher = {EDP Sciences},
}

@Article{Walmsley2019,
  author    = {Walmsley, Mike and Smith, Lewis and Lintott, Chris and Gal, Yarin and Bamford, Steven and Dickinson, Hugh and Fortson, Lucy and Kruk, Sandor and Masters, Karen and Scarlata, Claudia and Simmons, Brooke and Smethurst, Rebecca and Wright, Darryl},
  journal   = {MNRAS},
  title     = {Galaxy Zoo: probabilistic morphology through Bayesian CNNs and active learning},
  year      = {2019},
  issn      = {1365-2966},
  month     = oct,
  number    = {2},
  pages     = {1554--1574},
  volume    = {491},
  doi       = {10.1093/mnras/stz2816},
  publisher = {Oxford University Press (OUP)},
}

@Article{ElKholy2025,
  author    = {El-Kholy, R.I. and Hayman, Z.M.},
  journal   = {A\&A},
  title     = {Optimised sampling of SDSS-IV MaStar spectra for stellar classification using supervised models},
  year      = {2025},
  issn      = {1432-0746},
  month     = jan,
  doi       = {10.1051/0004-6361/202451309},
  publisher = {EDP Sciences},
}

@Article{Lochner2021,
  author    = {Lochner, M. and Bassett, B.A.},
  journal   = {Astron. Comput.},
  title     = {Astronomaly: Personalised active anomaly detection in astronomical data},
  year      = {2021},
  issn      = {2213-1337},
  month     = jul,
  pages     = {100481},
  volume    = {36},
  doi       = {10.1016/j.ascom.2021.100481},
  publisher = {Elsevier BV},
}

@Article{Quan2021,
  author    = {Quan, Yinghui and Zhong, Xian and Feng, Wei and Chan, Jonathan Cheung-Wai and Li, Qiang and Xing, Mengdao},
  journal   = {Remote Sens.},
  title     = {SMOTE-Based Weighted Deep Rotation Forest for the Imbalanced Hyperspectral Data Classification},
  year      = {2021},
  issn      = {2072-4292},
  month     = jan,
  number    = {3},
  pages     = {464},
  volume    = {13},
  doi       = {10.3390/rs13030464},
  publisher = {MDPI AG},
}

@Article{Yu2019,
  author    = {Yu, Hualong and Yang, Xibei and Zheng, Shang and Sun, Changyin},
  journal   = {IEEE Transactions on Neural Networks and Learning Systems},
  title     = {Active Learning From Imbalanced Data: A Solution of Online Weighted Extreme Learning Machine},
  year      = {2019},
  issn      = {2162-2388},
  month     = apr,
  number    = {4},
  pages     = {1088--1103},
  volume    = {30},
  doi       = {10.1109/tnnls.2018.2855446},
  publisher = {Institute of Electrical and Electronics Engineers (IEEE)},
}

@Article{Zhang2018,
  author    = {Zhang, Hang and Liu, Weike and Shan, Jicheng and Liu, Qingbao},
  journal   = {IEEE Access},
  title     = {Online Active Learning Paired Ensemble for Concept Drift and Class Imbalance},
  year      = {2018},
  issn      = {2169-3536},
  pages     = {73815--73828},
  volume    = {6},
  doi       = {10.1109/access.2018.2882872},
  publisher = {Institute of Electrical and Electronics Engineers (IEEE)},
}

@Article{Saha2018,
  author    = {Saha, S. and Basak, S. and Safonova, M. and Bora, K. and Agrawal, S. and Sarkar, P. and Murthy, J.},
  journal   = {Astron. Comput.},
  title     = {Theoretical validation of potential habitability via analytical and boosted tree methods: An optimistic study on recently discovered exoplanets},
  year      = {2018},
  issn      = {2213-1337},
  month     = apr,
  pages     = {141--150},
  volume    = {23},
  doi       = {10.1016/j.ascom.2018.03.003},
  publisher = {Elsevier BV},
}

@Article{Basak2021,
  author    = {Basak, Suryoday and Mathur, Archana and Theophilus, Abhijit Jeremiel and Deshpande, Gouri and Murthy, Jayant},
  journal   = {EPJ ST},
  title     = {Habitability classification of exoplanets: a machine learning insight},
  year      = {2021},
  issn      = {1951-6401},
  month     = jul,
  number    = {10},
  pages     = {2221--2251},
  volume    = {230},
  doi       = {10.1140/epjs/s11734-021-00203-z},
  publisher = {Springer Science and Business Media LLC},
}

@Article{Saha2020,
  author    = {Saha, Snehanshu and Nagaraj, Nithin and Mathur, Archana and Yedida, Rahul and H R, Sneha},
  journal   = {EPJ ST},
  title     = {Evolution of novel activation functions in neural network training for astronomy data: habitability classification of exoplanets},
  year      = {2020},
  issn      = {1951-6401},
  month     = nov,
  number    = {16},
  pages     = {2629--2738},
  volume    = {229},
  doi       = {10.1140/epjst/e2020-000098-9},
  publisher = {Springer Science and Business Media LLC},
}

@InBook{Patel2024,
  author    = {Patel, Yash},
  publisher = {IntechOpen},
  title     = {Predicting Exoplanets Habitability: Metrics and Models},
  year      = {2024},
  isbn      = {9781836343585},
  month     = dec,
  booktitle = {Anomaly Detection - Methods, Complexities and Applications},
  doi       = {10.5772/intechopen.1007686},
  issn      = {2633-1403},
}

@InProceedings{Dhama2025,
  author    = {Dhama, Ruthwik and Singh, Praveen Pratap and Kumar, Vishal},
  booktitle = {2025 8th International Conference on New Media Studies (CONMEDIA)},
  title     = {METHI: An Ensemble-based Machine Learned Exoplanetary Habitability Index},
  year      = {2025},
  month     = oct,
  pages     = {23--28},
  publisher = {IEEE},
  doi       = {10.1109/conmedia67140.2025.11290119},
}

@Article{RodriguezMozos2017,
  author    = {Rodríguez-Mozos, J. M. and Moya, A.},
  journal   = {MNRAS},
  title     = {Statistical-likelihood Exo-Planetary Habitability Index (SEPHI)},
  year      = {2017},
  issn      = {1365-2966},
  month     = aug,
  number    = {4},
  pages     = {4628--4636},
  volume    = {471},
  doi       = {10.1093/mnras/stx1910},
  publisher = {Oxford University Press (OUP)},
}

@Article{Barnes2015,
  author    = {Barnes, Rory and Meadows, Victoria S. and Evans, Nicole},
  journal   = {ApJ},
  title     = {COMPARATIVE HABITABILITY OF TRANSITING EXOPLANETS},
  year      = {2015},
  issn      = {1538-4357},
  month     = nov,
  number    = {2},
  pages     = {91},
  volume    = {814},
  doi       = {10.1088/0004-637x/814/2/91},
  publisher = {American Astronomical Society},
}

@Article{SchulzeMakuch2011,
  author    = {Schulze-Makuch, Dirk and Méndez, Abel and Fairén, Alberto G. and von Paris, Philip and Turse, Carol and Boyer, Grayson and Davila, Alfonso F. and António, Marina Resendes de Sousa and Catling, David and Irwin, Louis N.},
  journal   = {AsBio},
  title     = {A Two-Tiered Approach to Assessing the Habitability of Exoplanets},
  year      = {2011},
  issn      = {1557-8070},
  month     = dec,
  number    = {10},
  pages     = {1041--1052},
  volume    = {11},
  doi       = {10.1089/ast.2010.0592},
  publisher = {SAGE Publications},
}

@Article{Kopparapu2014,
  author    = {Kopparapu, Ravi Kumar and Ramirez, Ramses M. and SchottelKotte, James and Kasting, James F. and Domagal-Goldman, Shawn and Eymet, Vincent},
  journal   = {ApJ},
  title     = {HABITABLE ZONES AROUND MAIN-SEQUENCE STARS: DEPENDENCE ON PLANETARY MASS},
  year      = {2014},
  issn      = {2041-8213},
  month     = may,
  number    = {2},
  pages     = {L29},
  volume    = {787},
  doi       = {10.1088/2041-8205/787/2/l29},
  publisher = {American Astronomical Society},
}

@Article{Breiman2001,
  author    = {Breiman, Leo},
  journal   = {Mach. Learn.},
  title     = {Random Forests},
  year      = {2001},
  issn      = {1573-0565},
  month     = oct,
  number    = {1},
  pages     = {5--32},
  volume    = {45},
  doi       = {10.1023/a:1010933404324},
  publisher = {Springer Science and Business Media LLC},
}

@InBook{Bourlard1994,
  author    = {Bourlard, Hervé A. and Morgan, Nelson},
  pages     = {59--80},
  publisher = {Springer US},
  title     = {Multilayer Perceptrons},
  year      = {1994},
  isbn      = {9781461532101},
  booktitle = {Connectionist Speech Recognition},
  doi       = {10.1007/978-1-4615-3210-1_4},
}

@Book{Acquaviva2023,
  author    = {Acquaviva, Viviana},
  publisher = {Princeton University Press},
  title     = {Machine learning for physics and astronomy},
  year      = {2023},
  address   = {Princeton},
  isbn      = {9780691203928},
  note      = {Literaturverzeichnis: Seiten 249-256},
  pagetotal = {259},
  ppn_gvk   = {1848581882},
}

@Article{Andres2022,
  author    = {de Andres, Daniel and Yepes, Gustavo and Sembolini, Federico and Martínez-Muñoz, Gonzalo and Cui, Weiguang and Robledo, Francisco and Chuang, Chia-Hsun and Rasia, Elena},
  journal   = {MNRAS},
  title     = {Machine learning methods to estimate observational properties of galaxy clusters in large volume cosmological N-body simulations},
  year      = {2022},
  issn      = {1365-2966},
  month     = nov,
  number    = {1},
  pages     = {111--129},
  volume    = {518},
  doi       = {10.1093/mnras/stac3009},
  publisher = {Oxford University Press (OUP)},
}

@Article{Agarwal2023,
  author    = {Agarwal, A.},
  journal   = {ApJ},
  title     = {Classification of Blazar Candidates of Unknown Type in Fermi 4LAC by Unanimous Voting from Multiple Machine-learning Algorithms},
  year      = {2023},
  issn      = {1538-4357},
  month     = apr,
  number    = {2},
  pages     = {109},
  volume    = {946},
  doi       = {10.3847/1538-4357/acbdfa},
  publisher = {American Astronomical Society},
}

@Article{Zeraatgari2023,
  author    = {Zeraatgari, Fatemeh Zahra and Hafezianzadeh, Fatemeh and Zhang, Yanxia and Mei, Liquan and Ayubinia, Ashraf and Mosallanezhad, Amin and Zhang, Jingyi},
  journal   = {MNRAS},
  title     = {Machine learning-based photometric classification of galaxies, quasars, emission-line galaxies, and stars},
  year      = {2023},
  issn      = {1365-2966},
  month     = nov,
  number    = {3},
  pages     = {4677--4689},
  volume    = {527},
  doi       = {10.1093/mnras/stad3436},
  publisher = {Oxford University Press (OUP)},
}

@Article{Bhavanam2024,
  author    = {Bhavanam, Srinadh Reddy and Channappayya, Sumohana S. and P. K, Srijith and Desai, Shantanu},
  journal   = {Ap\&SS},
  title     = {Enhanced astronomical source classification with integration of attention mechanisms and vision transformers},
  year      = {2024},
  issn      = {1572-946X},
  month     = aug,
  number    = {8},
  volume    = {369},
  doi       = {10.1007/s10509-024-04357-9},
  publisher = {Springer Science and Business Media LLC},
}

@Article{Fluke2019,
  author    = {Fluke, Christopher J. and Jacobs, Colin},
  journal   = {WIREs DMKD},
  title     = {Surveying the reach and maturity of machine learning and artificial intelligence in astronomy},
  year      = {2019},
  issn      = {1942-4795},
  month     = dec,
  number    = {2},
  volume    = {10},
  doi       = {10.1002/widm.1349},
  publisher = {Wiley},
}

@Article{Cohn1996,
  author    = {Cohn, D. A. and Ghahramani, Z. and Jordan, M. I.},
  journal   = {JAIR},
  title     = {Active Learning with Statistical Models},
  year      = {1996},
  issn      = {1076-9757},
  month     = mar,
  pages     = {129--145},
  volume    = {4},
  doi       = {10.1613/jair.295},
  publisher = {AI Access Foundation},
}

@InBook{Balcan2007,
  author    = {Balcan, Maria-Florina and Broder, Andrei and Zhang, Tong},
  pages     = {35--50},
  publisher = {Springer Berlin Heidelberg},
  title     = {Margin Based Active Learning},
  year      = {2007},
  isbn      = {9783540729259},
  booktitle = {Learning Theory},
  doi       = {10.1007/978-3-540-72927-3_5},
}

@Article{Turnbull2015,
  author        = {Turnbull, Margaret C},
  title         = {ExoCat-1: The Nearby Stellar Systems Catalog for Exoplanet Imaging Missions},
  year          = {2015},
  month         = oct,
  archiveprefix = {arXiv},
  copyright     = {arXiv.org perpetual, non-exclusive license},
  doi           = {10.48550/ARXIV.1510.01731},
  eprint        = {1510.01731},
  primaryclass  = {astro-ph.SR},
  publisher     = {arXiv},
}

@Article{Stassun2019,
  author    = {Stassun, Keivan G. and Oelkers, Ryan J. and Paegert, Martin and Torres, Guillermo and Pepper, Joshua and Lee, Nathan De and Collins, Kevin and Latham, David W. and Muirhead, Philip S. and Chittidi, Jay and Rojas-Ayala, Bárbara and Fleming, Scott W. and Rose, Mark E. and Tenenbaum, Peter and Ting, Eric B. and Kane, Stephen R. and Barclay, Thomas and Bean, Jacob L. and Brassuer, C. E. and Charbonneau, David and Ge, Jian and Lissauer, Jack J. and Mann, Andrew W. and McLean, Brian and Mullally, Susan and Narita, Norio and Plavchan, Peter and Ricker, George R. and Sasselov, Dimitar and Seager, S. and Sharma, Sanjib and Shiao, Bernie and Sozzetti, Alessandro and Stello, Dennis and Vanderspek, Roland and Wallace, Geoff and Winn, Joshua N.},
  journal   = {AJ},
  title     = {The Revised TESS Input Catalog and Candidate Target List},
  year      = {2019},
  issn      = {1538-3881},
  month     = sep,
  number    = {4},
  pages     = {138},
  volume    = {158},
  doi       = {10.3847/1538-3881/ab3467},
  publisher = {American Astronomical Society},
}

@Article{Feng2017,
  author    = {Feng, F. and Tuomi, M. and Jones, H. R. A. and Barnes, J. and Anglada-Escudé, G. and Vogt, S. S. and Butler, R. P.},
  journal   = {AJ},
  title     = {Color Difference Makes a Difference: Four Planet Candidates around τ Ceti},
  year      = {2017},
  issn      = {1538-3881},
  month     = sep,
  number    = {4},
  pages     = {135},
  volume    = {154},
  doi       = {10.3847/1538-3881/aa83b4},
  publisher = {American Astronomical Society},
}

@Article{Lex2014,
  author    = {Lex, Alexander and Gehlenborg, Nils and Strobelt, Hendrik and Vuillemot, Romain and Pfister, Hanspeter},
  journal   = {IEEE Transactions on Visualization and Computer Graphics},
  title     = {UpSet: Visualization of Intersecting Sets},
  year      = {2014},
  issn      = {1077-2626},
  month     = dec,
  number    = {12},
  pages     = {1983--1992},
  volume    = {20},
  doi       = {10.1109/tvcg.2014.2346248},
  publisher = {Institute of Electrical and Electronics Engineers (IEEE)},
}

@Misc{Nothman2018,
  author       = {Joel Nothman},
  howpublished = {\url{https://github.com/jnothman/UpSetPlot}},
  title        = {upsetplot: A Python implementation of UpSet plots},
  year         = {2018},
}

@Article{Fawcett2006,
  author    = {Fawcett, Tom},
  journal   = {attern Recognit. Lett.},
  title     = {An introduction to ROC analysis},
  year      = {2006},
  issn      = {0167-8655},
  month     = jun,
  number    = {8},
  pages     = {861--874},
  volume    = {27},
  doi       = {10.1016/j.patrec.2005.10.010},
  publisher = {Elsevier BV},
}

@Article{Bergstra2012,
  author  = {James Bergstra and Yoshua Bengio},
  journal = {J. Mach. Learn. Res.},
  title   = {Random Search for Hyper-Parameter Optimization},
  year    = {2012},
  pages   = {281-305},
  volume  = {13},
  url     = {https://api.semanticscholar.org/CorpusID:15700257},
}

@Book{Huber1981,
  author    = {Huber, Peter J.},
  publisher = {Wiley},
  title     = {Robust Statistics},
  year      = {1981},
  isbn      = {9780471725251},
  month     = feb,
  doi       = {10.1002/0471725250},
  issn      = {1940-6347},
  journal   = {Wiley Series in Probability and Statistics},
}
\bibliographystyle{aasjournalv7}



\end{document}